\documentclass[sigconf, nonacm, pdfa]{acmart}

\newcommand\vldbdoi{XX.XX/XXX.XX}
\newcommand\vldbpages{XXX-XXX}
\newcommand\vldbvolume{17}
\newcommand\vldbissue{10}
\newcommand\vldbyear{2024}
\newcommand\vldbauthors{\authors}
\newcommand\vldbtitle{\shorttitle} 
\newcommand\vldbavailabilityurl{https://github.com/chaozhang-cs/bic}
\newcommand\vldbpagestyle{empty} 

\usepackage[ruled,linesnumbered]{algorithm2e}
\newcommand\window{\mathcal{W}}
\newcommand\rt[1]{find(#1)}
\newcommand\incre{partial}

\begin{document}

\title{Incremental Sliding Window Connectivity over Streaming Graphs\\ (Extended Version)\footnote{A shorter version of this paper has been accepted for publication in VLDB 2024.}}

\author{Chao Zhang}
\affiliation{%
  \institution{University of Waterloo}
  \city{Waterloo}
  \country{Canada}}
\email{chao.zhang@uwaterloo.ca}

\author{Angela Bonifati}
\affiliation{%
  \institution{Lyon 1 University, CNRS \& IUF}
  \city{Lyon}
  \country{France}}
\email{angela.bonifati@univ-lyon1.fr}

\author{M. Tamer Özsu}
\affiliation{%
  \institution{University of Waterloo}
  \city{Waterloo}
  \country{Canada}}
\email{tamer.ozsu@uwaterloo.ca}

\begin{abstract}
We study index-based processing for connectivity queries within sliding windows on streaming graphs.
These queries, which determine whether two vertices belong to the same connected component, are fundamental operations in real-time graph data processing and demand high throughput and low latency.
While indexing methods that leverage data structures for fully dynamic connectivity can facilitate efficient query processing, they encounter significant challenges with deleting expired edges from the window during window updates. 
We introduce a novel indexing approach that eliminates the need for physically performing edge deletions. 
This is achieved through a unique bidirectional incremental computation framework, referred to as the BIC model. The BIC model implements two distinct incremental computations to compute connected components within the window, operating along and against the timeline, respectively. These computations are then merged to efficiently compute queries in the window.
We propose techniques for optimized index storage, incremental index updates, and efficient query processing to improve BIC effectiveness. 
Empirically, BIC achieves a 14$\times$ increase in throughput and a reduction in P95 latency by up to 3900$\times$ when compared to state-of-the-art indexes.
\end{abstract}

\maketitle

\pagestyle{\vldbpagestyle}
\begingroup\small\noindent\raggedright
\textbf{PVLDB Reference Format:}\\
\vldbauthors. \vldbtitle. PVLDB, \vldbvolume(\vldbissue): \vldbpages, \vldbyear.\\
\href{https://doi.org/\vldbdoi}{doi:\vldbdoi}
\endgroup
\begingroup
\renewcommand\thefootnote{}\footnote{\noindent
%This work is licensed under the Creative Commons BY-NC-ND 4.0 International License. Visit \url{https://creativecommons.org/licenses/by-nc-nd/4.0/} to view a copy of this license. For any use beyond those covered by this license, obtain permission by emailing \href{mailto:info@vldb.org}{info@vldb.org}. Copyright is held by the owner/author(s). Publication rights licensed to the VLDB Endowment. \\
%\raggedright Proceedings of the VLDB Endowment, Vol. \vldbvolume, No. \vldbissue\ %
%ISSN 2150-8097. \\
%\href{https://doi.org/\vldbdoi}{doi:\vldbdoi} \\
}\addtocounter{footnote}{-1}\endgroup

\ifdefempty{\vldbavailabilityurl}{}{
\vspace{.3cm}
\begingroup\small\noindent\raggedright\textbf{PVLDB Artifact Availability:}\\
The source code, data, and/or other artifacts have been made available at \url{\vldbavailabilityurl}.
\endgroup
}
\setcounter{page}{1}

\section{introduction}\label{sec:intro}
Graphs have been the natural representation of data in many domains \cite{Newman2010,SakrBVIAAAABBDV21}, where individual entities are represented as vertices and the relationships between entities are represented as edges. 
With graph-structured data, one of the most interesting operations is to compute \textit{connected components} (CCs) \cite{10.1145/3186728.3164139,sahu2020ubiquity}, which are basically subsets of vertices in a graph such that all vertices in the subset are connected via undirected paths.
Analyzing CCs has many practical applications.
In social networks, CCs represent distinct communities or groups of individuals who are tightly connected to each other, and analyzing these components helps in identifying friend circles and influential users within the larger social network \cite{10.1177/016555150202800601}.
In transport networks, identifying CCs helps in understanding traffic flow and connectivity between different parts of a city, which is crucial for optimizing traffic signal timings, planning public transportation routes, and managing emergency response systems \cite{von2009public}.
In financial networks, CCs are used to detect unusual patterns or suspicious clusters of activity that differ from normal transaction patterns, which is crucial in fraud detection \cite{amy2023}.

In modern data-driven applications, stream processing \cite{10.1145/2588555.2595641,10.1145/2723372.2742788,10.14778/2536222.2536229,10.1145/2517349.2522737,carbone2015apache} is of significant importance,  providing real-time data processing capabilities. In the stream model, data arrive at a processing site continuously and each data record contains a payload and a timestamp.
Computations are typically performed over \textit{windows} that are continuous finite subsets of streaming data over the unbounded input stream \cite{10.1145/776985.776986}. Of particular interest are \textit{time-based sliding windows} that are characterized by a \textit{window size} and a \textit{slide interval}, denoted as $\alpha$ and $\beta$, respectively, which are given in time units. Each window contains data whose timestamps are within the window.
For instance, a sliding window with window size of $3$ hours and slide interval of $1$ hour includes all streaming data of the last $3$ hours, and the window is updated every hour by deleting expired data (\textit{i.e.}, data whose timestamp falls outside the window) from the window and inserting new streaming data.

In this paper, we study the computation of CCs over a time-based sliding window in \textit{a streaming graph} \cite{10.1145/2627692.2627694,10.1145/3318464.3389733} that is essentially a stream of edges. 
For ease of presentation, we focus on \textit{connectivity} queries that check whether two vertices belong to the same CC. 
Computing connectivity is equivalent to computing CCs as the former requires computing and storing CCs.
The problem of computing connectivity over sliding windows is referred to as \textit{sliding window connectivity}. 
Computing sliding window connectivity allows for the continuous analysis of data streams in real-time, enabling the immediate detection of changes or anomalies in network structures. This is crucial in scenarios like social network monitoring \cite{10.1177/016555150202800601}, traffic monitoring \cite{von2009public}, and fraud detection \cite{amy2023}, where timely responses are essential.

% We use the running example in Figure \ref{fig:sliding-window-connectivity} to explain the problem. This example will be used throughout the paper to explain our approach.

\begin{figure}
    \centering
    \resizebox{0.9\linewidth}{!}{
    \includegraphics{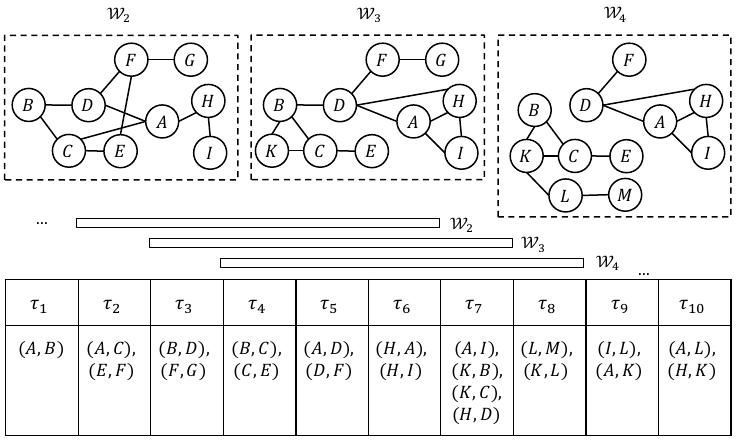}}
    \caption{Ruining example.}
    \label{fig:sliding-window-connectivity}
\end{figure}

\textit{Running example}.
Figure \ref{fig:sliding-window-connectivity} shows a query to compute CCs over a sliding window with window size of $5$  and  slide interval of $1$ (time units are not important). $\window_2$ is the instance of the sliding window ranging from timestamp $\tau_2$ to $\tau_6$, $\window_3$ is the one from $\tau_3$ to $\tau_7$, and so forth. 
$\window_2$ and $\window_3$ contain only one CC while $\window_4$ has two CCs. Vertices $C$ and $G$ are connected in both $\window_2$ and $\window_3$ but not in $\window_4$.

The naive approach to compute sliding window connectivity is to traverse the streaming graph in each window instance, \textit{e.g.}, performing depth-first-search (DFS) in each $\window_i$ in Figure \ref{fig:sliding-window-connectivity}.
The naive approach is to apply DFS to a streaming setting by performing it in each window. However, this approach would lead to recomputing CCs from scratch in each window before processing queries, thus being inefficient in a streaming setting.
A non-trivial method is to use data structures designed for fully dynamic connectivity (FDC) \cite{10.1145/945394.945398,10.1145/225058.225269,10.1145/276698.276715,10.1145/320211.320215,10.1145/265910.265914,10.1145/502090.502095,10.1145/335305.335345,10.1137/S0097539797327209,1.9781611973105.126,10.5555/2627817.2627898,kejlbergrasmussen_et_al:LIPIcs:2016:6395,10.5555/3039686.3039718,10.1145/3055399.3055415,10.1137/S0097539705447256,henzinger1998lower,10.14778/3551793.3551868,10.1145/335305.335345}.
Specifically, FDC supports $3$ operations: \texttt{insert}, \texttt{delete}, and \texttt{query}. 
As previously mentioned, when the sliding window needs to be updated, it is necessary to remove expired data while adding new data into the window, \textit{e.g.}, from $\window_2$ to $\window_3$ in Figure \ref{fig:sliding-window-connectivity}, it is necessary to delete streaming edges with timestamp $\tau_2$ and insert ones with timestamp $\tau_7$.
Obviously, the \texttt{insert}  and \texttt{delete} operations supported by FDC can be used to deal with the updates required by sliding windows.
% The main bottleneck of the FDC approach is that the \texttt{delete} operation can have high latency as it requires traversing the graph in the window in the worst case, which has the same cost as the naive approach.
The main performance issue of the FDC approaches is the \texttt{delete} operation. These use spanning trees to represent CCs, and deleting an edge of a spanning tree requires traversing the graph in the window to verify whether there exists an edge that can reconnect the two split sub-trees (or CCs). In the worst case, this takes the same time as the naive approach (see details in \S \ref{sec:related_works}). For example, from $\window_3$ to $\window_4$ in Figure \ref{fig:sliding-window-connectivity}, edge \textit{(B, D)} needs to be deleted from $\window_3$, which can lead to two CCs in the graph of $\window_3$. Then, it is necessary to traverse the entire graph of $\window_3$ to verify whether the two CCs can be reconnected.

In this paper, we propose the bidirectional incremental computation model (BIC) to process sliding window connectivity.
The main idea of BIC is that streaming edges with contiguous timestamps are grouped to form disjoint \textit{chunks}, windows are decomposed according to chunks, and queries are processed by applying partial computations in chunks followed by merging the corresponding partial results. 
Specifically, we compute two kinds of buffers for each chunk: \textit{forward} and \textit{backward} buffers.
Both forward and backward buffers are computed incrementally, achieved by scanning streaming edges in chunks. For the forward buffer, streaming edges are processed sequentially, starting from the first edge and progressing to the last edge within the chunk. Conversely, to compute the backward buffer, the streaming edges are scanned in the reverse order, beginning with the last edge and moving towards the first in the chunk.
These two kinds of buffers are stored and the elements in them are merged to compute the query result of each window. 

\begin{figure}
    \centering
    \resizebox{0.9\linewidth}{!}{
    \includegraphics{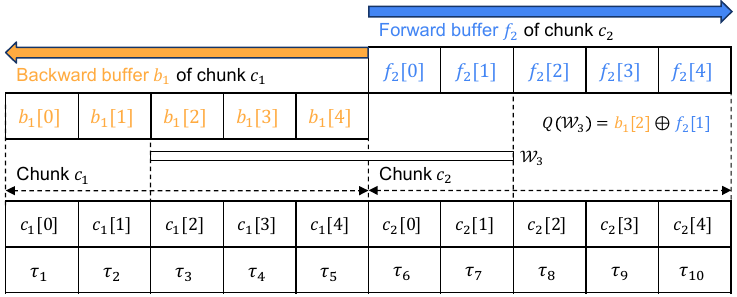}}
    \caption{
    % Running example of using the BIC model to compute sliding windows in Figure \ref{fig:sliding-window-connectivity}.
    Using BIC for the running example in Figure \ref{fig:sliding-window-connectivity}.
    }
    \label{fig:bic}
\end{figure}

Figure \ref{fig:bic} demonstrates the BIC model, using the example provided in Figure \ref{fig:sliding-window-connectivity}. In this example, each chunk contains streaming edges spanning $5$ timestamps. The forward buffer $f_2$ for chunk $c_2$ is computed by scanning streaming edges from timestamps $\tau_{6}$ to $\tau_{10}$. The backward buffer $b_1$ for chunk $c_1$ is computed by scanning streaming edges in reverse within $c_1$, from $\tau_{5}$ back to $\tau_{1}$.
Intuitively, $f_2[1]$ captures the connectivity information of edges spanning from $\tau_6$ to $\tau_7$, whereas $b_1[2]$ encompasses the connectivity information of edges from $\tau_3$ to $\tau_5$ (see \S \ref{sec:incremental_connectivity} for details). Consequently, any connectivity query $Q$ over $\window_3$ can be computed by merging these segments of connectivity information (see \S \ref{sec:merging_sub_connectivity} for details).
The novelty of the BIC model is its ability to deal with the deletion of expired streaming edges from the window without necessitating any corresponding deletions in the maintained index to reflect these changes.
For example, $Q$ over $\window_3$ is computed by merging the segments of connectivity information stored in $b_1[2]$ and $f_2[1]$, respectively. Our approach contrasts with a FDC approach, which necessitates deleting expired edges with $\tau_2$ from and inserting new edges with $\tau_7$ into the index of $W_2$.

In order to leverage the BIC model to compute sliding window connectivity, two main challenges need to be addressed.
Challenge C1 is related to the storage of backward buffers. 
Specifically, every element in a backward buffer needs to be stored, which will be retrieved for computing query results. 
However, storing all elements in a backward buffer will lead to significant overhead (see detail in \S \ref{sec:backward_storage}). 
Challenge C2 is on efficiently merging backward and forward buffers to compute query results. 
Vertices $s$ and $t$ might not be connected in a single buffer (either forward or backward one) but may be connected transitively via vertices that exist in both kinds of buffers. 
Such inter-buffer checking can result in searching entire buffers, which is not feasible to achieve low latency query processing (see detail in \S \ref{sec:merging_sub_connectivity}).


This paper makes the following contributions:
\begin{itemize}
    \item We introduce the BIC model, which transforms the updates of edge insertions and deletions in sliding window computations into bidirectional incremental computations involving only edge insertions. This approach effectively eliminates the need for the costly operation of deleting expired edges in maintaining indexes for processing connectivity queries in sliding windows.
    
    \item To tackle challenge C1, we propose an approach that stores just a single element in each backward buffer instead of all elements. The single element stored is capable of reconstructing all the other elements. This method introduces no additional overhead for query processing.

    \item To address challenge C2, we design a bipartite graph to bridge the global connectivity information between backward buffers and forward buffers, such that the search space of the merging operation can be significantly reduced.

    \item Our approach achieves near $O(\log{n})$ complexity for both worst-case query time and amortized index update time, with $n$ representing the number of vertices in a window. Although the number of CCs in the window influences these complexity results, it is noteworthy that the number of CCs is typically very small in real-world graphs.
    
    \item Comprehensive experimental evaluation using $6$ real-world datasets and $2$ synthetic datasets from industrial-grade benchmarks demonstrates that our approach achieves a 14$\times$ increase in throughput and a reduction in P95 latency by up to 3900$\times$ when compared to state-of-the-art indexes.
\end{itemize}
Due to space constraints, detailed proofs and algorithm pseudocodes are available in our online technical report \cite{zhang2024incremental}.
%We discuss related works in \S \ref{sec:related_works} and define the problem in \S \ref{sec:problem_statement}. 
%The BIC model is presented in \S \ref{sec:bic}, followed by approaches designed to address the three changes in \S \ref{sec:partial}, \S \ref{sec:merging}, respectively. We report the experimental results in \S \ref{sec:experiments} and conclude in \S \ref{sec:conclusion}.

\section{related work}\label{sec:related_works}
\textit{Fully dynamic connectivity}.
FDCs \cite{10.1145/945394.945398,10.1145/225058.225269,10.1145/276698.276715,10.1145/320211.320215,10.1145/265910.265914,10.1145/502090.502095,10.1145/335305.335345,10.1137/S0097539797327209,1.9781611973105.126,10.5555/2627817.2627898,kejlbergrasmussen_et_al:LIPIcs:2016:6395,10.5555/3039686.3039718,10.1145/3055399.3055415,10.1137/S0097539705447256,henzinger1998lower,10.14778/3551793.3551868,10.1145/335305.335345}
are data structures that can reflect the window updates due to expirations and insertions.
They support $3$ operations: \texttt{insert}, \texttt{delete}, and \texttt{query}.
Designing an algorithm that can perform all operations in $O(\log{n})$ time (even amortized time) is a long-standing problem. 
An early work shows the lower bound of $\Omega(\log{n}/\log{\log{n}})$ \cite{henzinger1998lower,miltersen1994complexity}, and the latest lower bound is proved to be $\Omega(\log{n})$ \cite{10.1137/S0097539705447256}.
The existing algorithms can be categorized into two classes: deterministic algorithms \cite{1.9781611973105.126,10.1145/276698.276715,10.1145/502090.502095,kejlbergrasmussen_et_al:LIPIcs:2016:6395,doi:10.1137/0214055,10.1145/265910.265914} and randomized algorithms \cite{10.1145/225058.225269,10.1145/320211.320215,10.5555/2627817.2627898,gibb2015dynamic,wang2015improved,10.1145/335305.335345,10.5555/3039686.3039718}. The former can return correct query result with a fixed time complexity while the latter has a random variable in either query results or time complexity.
There exist two sub-classes in the randomized ones: Monte Carlo algorithms \cite{10.5555/2627817.2627898,gibb2015dynamic,wang2015improved} and Las Vegas algorithms \cite{10.1145/225058.225269,10.1145/320211.320215,10.1145/335305.335345,10.5555/3039686.3039718}. Monte Carlo algorithms have a fixed time complexity but the query results might be incorrect, while Las Vegas algorithms can return correct query results but the time complexity contains a random variable.
Deterministic algorithms can be further classified according to their primary focus: amortized algorithms \cite{1.9781611973105.126,10.1145/276698.276715,10.1145/502090.502095} and worst-case algorithms \cite{kejlbergrasmussen_et_al:LIPIcs:2016:6395,doi:10.1137/0214055,10.1145/265910.265914}, which are designed to optimize amortized and worst-case time complexity, respectively. 
According to the latest study \cite{10.14778/3551793.3551868}, most of these algorithms rely on complicated data structures, which make them hard to implement.
The seminal works, HK \cite{10.1145/225058.225269,10.1145/320211.320215} and HDT \cite{10.1145/276698.276715,10.1145/502090.502095}, are the only two algorithms that have been successfully implemented and compared \cite{10.1145/264216.264223,10.1145/945394.945398}.
HK is a Las Vegas randomized algorithm while HDT is a deterministic algorithm to amortize the time complexity. 
Both algorithms are designed based on a framework using spanning trees and incorporate specific techniques to deal with the main bottleneck in the framework. We briefly discuss the framework below.
The overarching idea is to use Euler-Tour Tree \cite{10.1145/225058.225269} to store the spanning trees of the input graph. 
Updates related to non-tree edges are trivial as the connected components will not be changed.
Tree edge insertions can be addressed by using the combine operation provided by Euler-Tour Tree.
The main problem is the case of tree edge deletion because it requires splitting the spanning tree into two sub-trees and then checking whether there exists a non-tree edge that can reconnect the two sub-trees. 
Such a non-tree edge is known as a \textit{replacement edge}.
Searching for a replacement edge requires traversing the entire graph in the worst case, which takes $O(|V|+|E|)$ time if BFS or DFS is used.
HK and HDT design advanced techniques to amortize the cost of searching for replacement edges. 
D-Tree \cite{10.14778/3551793.3551868} is a recent work that also uses the spanning tree framework, but includes a different design to deal with the problem. Specifically, D-Tree balances the length from each vertex to its root so as to reduce the average cost of searching. 
However, the worst-case time complexity of all existing approaches remains the same as using BFS or DFS.
In this paper, we focus on designing a deterministic algorithm for computing connectivity queries within sliding windows. 
Thus, we adopt D-Tree as the current state-of-the-art FDC approach. 
We also include HDT as a baseline in our experimental evaluation as HDT can be faster than HK according to early experiments \cite{10.1145/264216.264223} and D-Tree is shown to be superior to HK.

\textit{Incremental connectivity}. When the input graph only has edge insertions, the well-known Union-Find (UF) \cite{10.1145/321879.321884,TARJAN1979110} can be used to compute connectivity queries, which supports two operations: \texttt{insert} and \texttt{query} in $O(\log{n})$ time, where $n$ is the number of vertices in the graph. 
However, UF cannot deal with edge deletions that are necessary in computing sliding window connectivity. In this paper, we design the bidirectional computation model to completely avoid edge deletions, such that UF  can be adopted. 

% \textit{Reachability and regular path queries}.
% Reachability queries check whether there is a directed path from a source vertex to a target vertex in a directed graph.
% The data structures that can support efficient reachability query processing are known as \textit{reachability indexes} (see \cite{yu2010graph,10.5555/3307192,10.1145/3555041.3589408,zhang2023indexing} for literature review).
% Computing reachability indexes is very expensive due to the underlying computing and compressing transitive closure of the input graph.
% In addition, reachability queries are completely different from connectivity queries as the latter does not take the direction of edges in paths into account.
% Thus, it is not feasible to use reachability indexes to process connectivity queries, not to mention that most reachability indexes are designed for static graphs.
% Recent works \cite{10.1145/3318464.3389733,9835463} study processing regular path queries (RPQs) in sliding windows over streaming graphs. 
% RPQs essentially check the reachability according to a path constraint defined as a regular expression using edge labels in an edge-labeled graph. 
% The overarching idea is to maintain a reachability index for the streaming graph in the window that can satisfy the defined path constraint, and the reachability index is continuously updated.
% Therefore, these approaches are not usable to computing sliding window connectivity due the same reason discussed above.

\textit{Stream processing systems (SPSs)}.
One of the primary requirements in general SPSs \cite{10.1145/2588555.2595641,10.1145/2723372.2742788,10.14778/2536222.2536229,10.1145/2517349.2522737,carbone2015apache} is to achieve high-throughput and low-latency computations. 
Two general stream processing models exist: \textit{micro-batch model}, \textit{e.g.}, Apache Spark \cite{10.1145/2517349.2522737}, and \textit{continuous model}, \textit{e.g.}, Apache Flink \cite{carbone2015apache}. The former starts the computation when each fix-sized batch is full while the latter immediately processes streaming data. 
In this paper, we adopt the continuous model, where the query result within a window is computed immediately upon the arrival of a streaming edge that completes the window.
This approach facilitates low-latency real-time computation, a key advantage of the continuous model \cite{carbone2015apache}.
There also exist graph streaming systems designed for graph analytics \cite{10.1145/3302424.3303974,10.1145/3267809.3267811,10.1145/2960414.2960419,10.1007/978-3-319-43659-3_24,10.1145/3364180,6408680} and SPARQL query processing \cite{10.1145/1526709.1526856,10.1007/978-3-642-25073-6_24,10.1007/978-3-642-17746-0_7,10.1007/978-3-319-25639-9_48}, and systems for temporary graph analytics \cite{10.1007/s00778-021-00667-4}. 
However, these systems lack specific designs for sliding window connectivity queries, and our approach proposed in the paper can be adopted into these systems to support such fundamental operations in graph computation.

\section{problem statement}\label{sec:problem_statement}
\textit{Connectivity queries in graphs}.
We denote an (undirected) graph as $G=(V,E)$, where $V$ is a finite set of vertices and $E\subseteq V \times V$ is a finite set of (undirected) edges.
A connectivity query $Q_c(s,t)$ in $G$ checks whether vertices $s$ and $t$ in $G$ are connected, and $Q_c(s,t)=True$ if there exists a path of undirected edges from $s$ to $t$ in $G$.

\textit{Streaming graphs}.
A streaming edge is an undirected edge with a timestamp, denoted as $e=(u,v,\tau)$, where $u$ and $v$ are the endpoints of $e$, and $\tau$ is the timestamp of $e$.
A streaming graph is an infinite sequence of streaming edges $(e_1,e_2,...)$, where subscript $i$ of $e_i$ denotes the arrival order of $e_i$, which is strictly increasing, and $e_i.\tau \leq e_j.\tau$ for any pair of streaming edges $e_i $ and $e_j$ such that $i<j$.

\textit{Streaming graphs in sliding windows}.
In stream processing, the computations are done in windows as the input stream is infinite. 
A window is a fixed-size subsequence of the input stream, denoted as $\window$.
In streaming graph processing, windows are usually \textit{time-based} that have fixed time intervals. 
A time-based window $\window$ has a beginning timestamp $\window.\tau_b$ and an ending timestamp $\window.\tau_e$.
$\window$ over a streaming graph $SG=(e_1,e_2,...)$ consists of all the streaming edges $e_i$ such that $\window.\tau_b \leq e_i.\tau \leq \window.\tau_e$.
% , \textit{i.e.}, $\window(SG)=\{e_i| \forall e_i\in SG, e_i.\tau \in [ \window.\tau_b,\window.\tau_e]\}$.
A sliding window is defined with a \textit{window size} $\alpha$ and a \textit{slide interval} $\beta$ given in time units, which essentially define the following sequence of window instances, \textit{i.e.}, $(\window_1,\window_2,...)$, such that for each $\window_i$, $\window_i.\tau_e = \window_i.\tau_b + \alpha$, and for every two adjacent $\window_i$ and $\window_{i+1}$, 
we have $\window_{i+1}.\tau_b = \window_{i}.\tau_b + \beta$.
% We note that $\alpha$ is larger than $\beta$, \textit{i.e.}, window instances have overlapping timestamps.

\textit{Sliding window connectivity}.
Sliding window connectivity computes connectivity queries 
 in each window instance of a sliding window. 
 We formalize the problem below.
 \begin{definition}[\textbf{Sliding Window Connectivity}]
     Given a streaming graph $SG$ and a time-based sliding window $\window(SG)$ defined by window size $\alpha$ and slide interval $\beta$, sliding window connectivity is to compute connectivity query $Q_c(s,t)$ between vertices $s$ and $t$ in all window instances $(\window_1,\window_2,...)$ of $\window(SG)$.
 \end{definition}

\begin{example}
Figure \ref{fig:sliding-window-connectivity} includes three window instances $\window_2$, $\window_3$, and $\window_4$ of the running example. 
For query $Q_c(C,G)$, a path exists between $C$ and $G$ in $\window_2$ and $\window_3$, making $Q_c(C,G)=True$, but in $\window_4$, $C$ and $G$ are not connected, so $Q_c(C,G)=False$.   
\end{example}

We aim to build an index for efficiently processing connectivity queries within sliding windows due to the inefficiencies of the naive approach based on graph traversal.
The main challenge is maintaining the index on the fly. Specifically, when the window is sliding, \textit{expired edges} will be deleted and \textit{new edges} will be inserted.
For instance, in Figure \ref{fig:sliding-window-connectivity},  when $\window_3$ is sliding to $\window_4$, \textit{expired edges} $(B,D)$ and $(F,G)$ with timestamp $\tau_3$ are deleted and \textit{new edges} $(L,M)$ and $(K,L)$ with timestamp $\tau_8$ are inserted.
The gist of the problem, therefore, is to effectively manage graph updates so they can be efficiently incorporated into an index, which in turn facilitates efficient processing of queries.

\section{A bidirectional incremental model}\label{sec:bic}
In this section, we propose the bidirectional incremental computation (BIC) model to address the problem of dynamic computations in sliding windows over streaming graphs. 
BIC's main idea is to group streaming edges into \textit{chunks} and perform \textit{forward} and \textit{backward} computations in chunks.
Both forward and backward computations can be done incrementally.
The final computation result in each window instance $\window_i$ can be obtained by merging sub-computation results obtained by the forward and backward computations.
We explain the main BIC concepts below, where we use the running example in Figure \ref{fig:bic} to guide the presentation.

In the BIC model, the result of a query $Q$ in each window instance $\window$ is computed as follows:
\begin{equation}\label{equ:partial_and_merging}
  Q(\window) = \incre(\window^1) \oplus \incre(\window^2),  
\end{equation}
where $\window^1$ and $\window^2$ are disjoint sub-windows of $\window$, \textit{i.e.}, $\window = \window^1 \cup \window^2$, and $\incre()$ and $\oplus$ are the partial and merging operations for processing $Q$, respectively.
$\incre()$ and $\oplus$ will be detailed in \S \ref{sec:partial} and \S \ref{sec:merging}, respectively. 

\begin{definition}[Chunks]
Given a streaming graph $SG$, streaming edges in $SG$ are grouped into non-overlapping \textit{chunks} $(c_1,c_2,..., c_n)$.
Each chunk $c$ is an array of slide intervals. The number of slide intervals in each chunk is the chunk size, referred to as $|c|$. 
\end{definition}

\begin{example}
For the running example in Figure \ref{fig:sliding-window-connectivity}, we present (Figure \ref{fig:bic}) the BIC model with chunk size of $5$ slide intervals.
Chunks $c_1$ and $c_2$ are arrays of $5$ elements, including $(c_1[0],...,c_1[4])$ and $(c_2[0],...,c_2[4])$, respectively.
Window instance $\window_3$ is split into two sub-windows $\window^1_3=\{\tau_3,\tau_4,\tau_5\}$ and $\window^2_3=\{\tau_6,\tau_7\}$.
Computing query $Q(\window_3)$ can be done by first computing $\incre(\window^1_3)$ and $\incre(\window^2_3)$, followed by merging the two partial results.    
\end{example}

\begin{definition}[Backward and Forward Buffers]
Given a streaming graph $SG$ and chunks $(c_1,c_2,...,c_n)$ over $SG$.
Two kinds of buffers of size $|c|$ are computed in each chunk $c_i$.
\begin{itemize}
    \item \textit{Forward buffer} $f_i$ consists of $|c|$ elements $(f_i[0],...,f_i[|c|-1])$, and each element $f_i[j]=\incre(c_i[0] \cup ... \cup c_i[j])$;
    \item \textit{Backward buffer} $b_i$ consists of $|c|$ elements $(b_i[0],...,b_i[|c|-1])$, and each element
    $b_i[j]=\incre(c_i[j]\cup ... \cup c_i[|c|-1])$.
\end{itemize}
\end{definition}

In Figure \ref{fig:bic}, $b_1[2]$ and $f_2[1]$ are computed as follows: $b_1[2]=\incre(c_1[2] \cup c_1[3]\cup c_1[4])$; $f_2[1]=\incre(c_2[0]\cup c_2[1])$.
Note that each chunk has a backward and a forward buffer, and we only show $b_1$ and $f_2$ in Figure \ref{fig:bic} for ease of presentation.

In this paper, we focus on using the chunk size that matches the window size divided by the slide interval, facilitating the merging of single elements from backward and forward buffers, respectively.

\begin{definition}[The BIC Model]
Given a query $Q$ and a sliding window over streaming graph $SG$, the BIC model computes a forward buffer $f_i$ and a backward buffer $b_i$ for each chunk $c_i$ in $SG$.
The computation of $Q$ in each window instance $\window$ is decomposed into 
partial computations of $Q$ over sub-windows: $Q(\window) = \incre(\window^1) \oplus \incre(\window^2)$, where $\incre(\window^1)$ and $ \incre(\window^2)$ are obtained by using $b_i$ and $f_i$.
\end{definition}

\begin{example}\label{example:bic-model}
For the running example in Figure \ref{fig:bic},  we have  $Q(\window_3)=b_1[2]\oplus f_1[2]$ because  
$Q(\window_3)$ is split into sub-windows  $\window_3^1$ and $\window_3^2$,
and  $\incre(\window_3^1)=b_1[2]$ and $\incre(\window_3^2)=f_2[1]$.   
\end{example}

For each chunk $c_i$, $f_i$, and  $b_i$  can be computed incrementally as long as the partial operation has the following property: 
\begin{equation}\label{equ:distributive_partial}
\incre(e_1\cup ... \cup e_{m})=\incre(\incre(e_1\cup ... \cup e_{m-1})\cup e_{m}).
\end{equation}
The partial operation that has the above property for computing connectivity queries is called \textit{incremental connectivity} \cite{10.1145/321879.321884,TARJAN1979110}, which will be detailed in \S \ref{sec:partial}.

The remainder of the paper will address the problem of  deploying sliding window connectivity $Q_c$ into the BIC model. 
In particular, we design $partial()$ in \S \ref{sec:partial} and $\oplus$ in \S \ref{sec:merging} for $Q_c$ such that Equations (\ref{equ:partial_and_merging}) and (\ref{equ:distributive_partial}) in the BIC model can be satisfied. 
\begin{figure*}
\begin{minipage}{0.463\linewidth}
    \centering
    \resizebox{\linewidth}{!}{
    \includegraphics{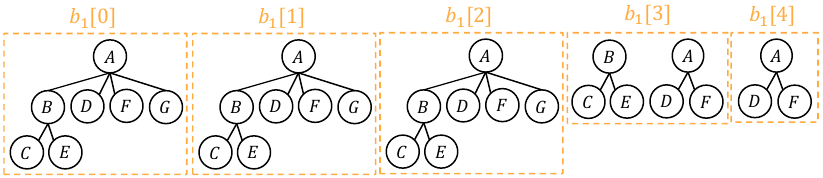}}
\end{minipage}
\hfill
\begin{minipage}{0.533\linewidth}
    \centering
    \resizebox{\linewidth}{!}{
    \includegraphics{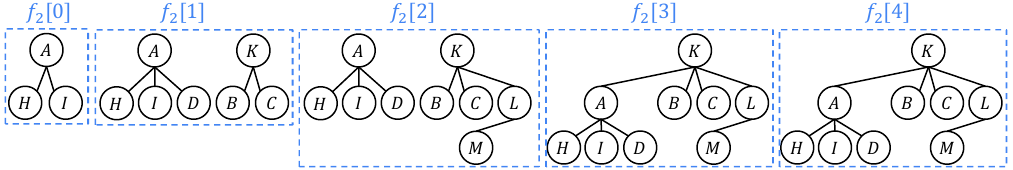}}
\end{minipage}
        \caption{The forward buffer $f_2$ over chunk $c_2$ and the backward buffer $b_1$ over chunk $c_1$ in the running example in Figure \ref{fig:bic}.}
    \label{fig:backward_forward_buffer}
\end{figure*}

\section{Incremental sub-connectivity}\label{sec:partial}
In this section, we discuss the design of $partial()$ for computing $Q_c$ over sliding windows in BIC, such that Equations (\ref{equ:partial_and_merging}) and (\ref{equ:distributive_partial}) are satisfied. 
Intuitively, the Union-Find Tree (UFT) \cite{10.1145/321879.321884,TARJAN1979110} algorithm can be applied to incrementally compute connectivity in backward and forward buffers. 
We identify the underlying challenges on buffer storage in \S \ref{sec:buffer_challenge}  and propose a snapshot-based approach to address the challenges in \S \ref{sec:backward_storage}.

\subsection{Incremental sub-connectivity in buffers}\label{sec:incremental_connectivity}
Our main finding is that \textit{incremental connectivity can be the partial operation for computing $Q_c$ in the BIC model, which allows fully incremental computations in the backward and forward buffers}.
Incremental connectivity is the case of computing connectivity queries over a dynamic graph with only edge insertions, which indeed satisfies the property required in Equation (\ref{equ:distributive_partial}).
Incremental connectivity can be efficiently computed using UFTs. 
Thus, in order to compute $Q_c$ using the BIC model, each buffer (either $f$ or $b$) computes UFTs to support the partial operation. 
We briefly review how UFT operates with respect to edge insertions and use the running example to explain the data structures in buffers.

Each UFT represents a CC in the graph.
UFTs are equipped with two operations: \textit{find} and \textit{union}.
The \textit{find} operation computes the root of a vertex in the UFT, and the \textit{union} operation links the two roots by making one of them a child of the other.
A connectivity query between $s$ and $t$ can be simply computed by checking whether $\rt{s}=\rt{t}$.
When an edge $(u,v)$ is inserted into the graph, if $\rt{u}=\rt{v}$, no update in the UFT is required since $u$ and $v$ are already in the same CC. 
Otherwise, the \textit{union} operation is performed to link the two UFTs with roots $\rt{u}$ and $\rt{v}$, respectively.
In this case, $u$ and $v$ are  connected by having $\rt{u}$ as a child of $\rt{v}$, or vice versa.
% , aka \textit{path compression} \cite{0029345}. 

In BIC, we compute UFTs in each forward buffer $f_i$ of chunk $c_i$, which are computed by continuously inserting streaming edges in $(c_i[0],...,c_i[|c|-1])$. 
Each $f_i[j]$ corresponds to the \textit{snapshot} of the UFTs after inserting streaming edges in $c_i[j]$.
The computation of the backward buffer $b_i$ is the same as the computation of $f_i$ except that UFTs in $b_i$ are computed by inserting streaming edges in the order of $(c_i[|c|-1], ..., c_i[0])$. 
Hereafter, we use the term snapshot to denote $f_i[j]$ or $b_i[j]$.

\begin{example}
Consider running the example in Figures \ref{fig:sliding-window-connectivity}  and  \ref{fig:bic}. 
The corresponding forward buffer $f_2$ and backward buffer $b_1$ are presented in Figure \ref{fig:backward_forward_buffer}.
$f_2$ is computed by inserting streaming edges in chunk $c_2$ from $c_2[0]$ to $c_2[4]$ into the UFTs of $f_2$, with its UFT snapshots illustrated in Figure \ref{fig:backward_forward_buffer}.
% Two situations may arise when inserting an edge into $f_2$: \textit{vertices connected}, \textit{e.g.}, the edge $(A,I)$ in $c_2[1]$ inserted into $f_2[0]$ does not alter the UFT because $A$ and $I$ are already connected, sharing the same root; \textit{vertices unconnected}, \textit{e.g.}, inserting edge $(I,L)$ from $c_2[3]$ into $f_2[2]$ necessitates a \textit{union} operation since $I$ and $L$ have different roots in $f_2[2]$. $A$ is linked as a child of $K$ as the UFT rooted at $A$ has fewer vertices than the UFT rooted at $K$ (an optimization technique explained later).
% 
Two situations may arise when inserting an edge into $f_2$.
The first situation is that vertices are already connected.  
For instance, the edge $(A,I)$ in $c_2[1]$ inserted into $f_2[0]$ does not alter the UFT because $A$ and $I$ are already connected,  sharing the same root.
The second situation is that vertices are not connected, so \textit{union} needs to be performed.
For instance, inserting edge $(I,L)$ from $c_2[3]$ into $f_2[2]$ necessitates a \textit{union} operation since $I$ and $L$ have different roots in $f_2[2]$.
$A$ is linked as a child of $K$ as the UFT rooted at $A$ has fewer vertices than the UFT rooted at $K$ (an optimization technique explained later).
The computation of $b_1$ works similarly but scans chunk $c_1$ in reverse, from $c_1[4]$ to $c_1[0]$.
$b_1[3]$ shows the snapshot after inserting edges from $c_1[4]$ and $c_1[3]$. 
\end{example}

In UFTs, \textit{find} is the building block for query processing and edge insertions. Thus, its cost should be reduced. We define  an optimization technique on UFTs in backward and forward buffers.

\begin{definition}[Optimized UFT]
If \textit{union} always makes the root of the smaller UFT as a child of the larger UFT, then the resulting UFT is an optimized UFT, where the size of a UFT ($|UFT|$) is the number of vertices in it.
\end{definition}
\begin{lemma}\label{lemma:optimized_uft}
The worst-case time complexity of performing find in an optimized UFT is $O(\log(|UFT|))$.
\end{lemma}
The lemma can be proved by induction \cite{0029345}.
In the remainder of the paper, any UFT in the BIC model is an optimized UFT. 
By abuse of notation, we simply denote optimized UFTs as UFTs.

\subsection{Computing and accessing buffers}\label{sec:buffer_challenge}
% We discuss the challenge in computing and storing the forward and backward buffers in the BIC model.
% To understand the challenge, we first define the computing and accessing order.
% Given a buffer, the order in which the snapshots in the buffer are computed and accessed are denoted as the \textit{computing order} and the \textit{accessing order} of the buffer, respectively.
% The computing order of the buffer is \textit{forward} (or \textit{backward}) if the snapshots in the buffer are computed sequentially from the first to the last (or from the last to the first).
% Similarly, forward (or backward) accessing order is the case that snapshots in the buffer are accessed from the first to the last (or from the last to the first).
% According to the BIC model, we have: 
% forward buffers have the forward computing order and the forward accessing order while backward buffers have the backward computing order but the forward accessing order.
We explore the challenges of computing and storing forward and backward buffers in  BIC, starting with the definition of computing and accessing orders.
The computing order is forward if snapshots are computed from the first to the last, and backward if from the last to the first. Accessing order follows the same logic; forward for accessing from the first to the last snapshot, and backward for the reverse. Thus, forward buffers have both forward computing and accessing orders, while backward buffers have backward computing and forward accessing orders.

\begin{example}
Consider the running example in Figures \ref{fig:bic} and \ref{fig:backward_forward_buffer}.
The computations of $Q_c$ in window instances $\window_2$, $\window_3$, and $\window_4$ using $b_1$ and $f_2$ are shown as follows: 
$Q_c(\window_2) = b_1[1]\oplus f_2[0]$; $Q_c(\window_3) = b_1[2]\oplus f_2[1]$; $Q_c(\window_4) = b_1[3]\oplus f_2[2]$.
% \begin{itemize}
%     \item $Q_c(\window_2) = b_1[1]\oplus f_2[0]$;
%     \item $Q_c(\window_3) = b_1[2]\oplus f_2[1]$;
%     \item $Q_c(\window_4) = b_1[3]\oplus f_2[2]$.
% \end{itemize}
Both $b_1$ and $f_2$ are accessed in the forward manner.
However, $b_1$ is computed in the backward manner while $f_2$ is computed in the forward manner.
\end{example}

To sum up, snapshots in forward buffers are first-computed-first-accessed (FCFA) while snapshots in backward buffers are first-computed-last-accessed (FCLA). 
FCFA is trivial to address, but FCLA will bring challenges in storing, discussed in \S \ref{sec:backward_storage}.

\subsection{Backward buffer storage}\label{sec:backward_storage}
Since forward buffer snapshots are FCFA, they can be processed and accessed immediately. For instance, in Figure \ref{fig:bic} at timestamp $\tau_8$, only $f_2[2]$ needs storage; $f_2[0]$ and $f_2[1]$ are unnecessary. 
Conversely, backward buffers are FCLA, needing more storage; for example, $b_1[3]$ is computed before and accessed after $b_1[1]$, thus must be stored during $b_1[1]$'s access.

The naive solution to deal with FCLA in backward buffer $b$ is to record every snapshot in $b$.
The corresponding storage cost is unfeasibly high because the size of UFTs in $b$ can be very large, considering that each chunk can have millions of streaming edges, and we have to make $|c|$ copies of each UFT.
To address the issue, we design a snapshot isolation approach.
We denote the edges in UFTs as Union-Find Tree Edges (UFTEs). 
Notice that UFTEs are not necessarily the same as streaming edges.
Our approach labels each UFTE with a snapshot index in $b$ to mark when the UFTE is inserted into $b$.
Consequently, $b_i[j]$ becomes a subset of UFTEs in $b[0]$, with each UFTE's snapshot index equal to or great than $j$. 
% Then, $b_i[j]$ is essentially a subset of UFTEs in $b[0]$ such that the snapshot index of each UFTE is not smaller than $j$.

\begin{figure}[h]
    \centering
    \resizebox{\linewidth}{!}{
    \includegraphics{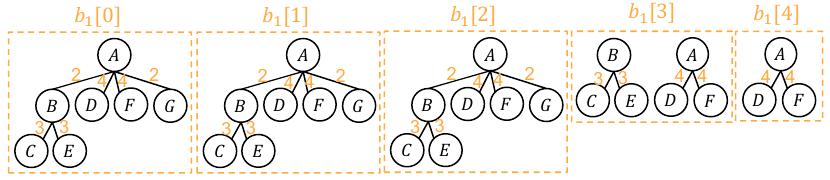}}
    \caption{Storing $b_1$ in Figure \ref{fig:backward_forward_buffer} using snapshot isolation.}
    \label{fig:backward_buffer_labeling}
\end{figure}
\begin{example}    
Figure \ref{fig:backward_buffer_labeling}  illustrates the snapshot isolation approach, using the running example in Figure \ref{fig:backward_forward_buffer}. UFTEs $(A,D)$ and $(A,F)$ are inserted into $b_1[4]$ and labeled with snapshot index $4$. Similarly, other UFTEs are labeled accordingly. To compute $b_1[3]$ from $b_1[0]$, we retrieve UFTEs in $b_1[0]$ labeled with snapshot indexes $3$ and $4$, specifically $(A,D)$, $(A,F)$, $(B,C)$, and $(B,E)$.
    % We present in Figure \ref{fig:backward_buffer_labeling} the snapshot isolation approach using the running example in Figure \ref{fig:backward_forward_buffer}.
    % UFTEs $(A,D)$ and $(A,F)$ are inserted into $b_1[4]$, such that they are labeled with snapshot index $4$, and the other UFTEs are labeled in the same way. 
    % In order to compute $b_1[3]$ from $b_1[1]$, we only need to retrieve UFTEs from $b_1[1]$ that are labeled with snapshot index $3$ and $4$, \textit{i.e.}, $(A,D)$, $(A,F)$, $(B,C)$, and $(B,E)$.
\end{example}

Lemma \ref{lemma:snapshot_isolation} shows that the \textit{find} operation can be correctly computed by using snapshot isolation to store backward buffer $b$.
% \revise{The proof is included in Appendix \ref{app:lemma} in the supplementary material.}

\begin{lemma}\label{lemma:snapshot_isolation}
Given vertex $v$ in $b_i[j]$, computing $\rt{v}$ in $b_i[j]$ is equivalent to computing $\rt{v}$ in $b_i[j'], j'<j$, if the traversal from $v$ in $b_i[j']$ terminates whenever a UFTE labeled with snapshot index that is smaller than $j$ is visited. 
\end{lemma}

Lemma \ref{lemma:snapshot_isolation} can be simply transformed into an algorithm for computing \textit{find} in UFTs labeled with snapshot indexes, which has the same time complexity $O(\log{|UFT|})$ as the original \textit{find} operation. 
% \revise{We present the algorithm in Algorithm \ref{algo:snapshot_isolation} in Appendix \ref{app:snapshot} in the supplementary material.}   

In Figure \ref{fig:backward_forward_buffer}, $\rt{C}$ in $b_1[3]$ is $B$.
If we perform the same computation  using $b_1[1]$ in Figure \ref{fig:backward_buffer_labeling}, UFTE $(B,C)$ will be visited as it is labeled with $3$. 
However, the next UFTE $(A,B)$ will not be visited as it is labeled with $2$ which means $(A,B)$ does not exist in $b_1[3]$. 
Then, \textit{find} stops and $B$ is returned as the root of $C$.

% With Lemma \ref{lemma:snapshot_isolation}, we only need to store $b_i[1]$ as the \textit{find} operation in any $b_i[j], j>1$ can be correctly computed
% ($b_i[0]$ is not needed as it contains the same information as $f_i[0]$).
% Consequently, the snapshot isolation approach only requires $O(|UFT|)$ space, compared to  $O(|UFT||c|)$ in the naive approach.
% For storing $b_1$ in the example of Figure 3, the naive approach needs to store $24$ UFTEs while snapshot isolation only needs to store $6$ UFTEs labeled with $6$ integers.
Lemma \ref{lemma:snapshot_isolation} allows us to store only $b_i[1]$ as the \textit{find} operation in any $b_i[j], j>1$ can be correctly computed
($b_i[0]$ is not needed as it contains the same information as $f_i[0]$).
Therefore, snapshot isolation necessitates only $O(|UFT|)$ space, compared to  $O(|UFT||c|)$ in the naive approach.
For storing $b_1$ in the example of Figure 3, the naive approach needs to store $24$ UFTEs while snapshot isolation only needs to store $6$ UFTEs labeled with $6$ integers.

\section{merging sub-connectivity}\label{sec:merging}
In this section, we propose the merging operation $\oplus$ for computing sliding window connectivity using the BIC model.
The primary goal is to merge the sub-connectivity information over sub-windows stored in backward and forward buffers for computing the query result in a full window. 
We discuss the challenges of merging, followed by a tailored data structure for efficient merging.

\subsection{Merging sub-connectivity}\label{sec:merging_sub_connectivity}
In the BIC model, any query $Q_c$ over each window $\window$ is computed as follows
$Q_c(\window)=partial(\window^1)\oplus partial(\window^2)$, 
where $partial(\window^1)$ and $partial(\window^2)$ are stored in $b_i[j]$ and $f_{i+1}[j-1]$, and $\oplus$ is the operation for merging these two partial results.
Given $Q_c(s,t)$, the main idea of merging is that if $s$ and $t$ are connected in either $b_i[j]$ or $f_{i+1}[j-1]$, then $Q_c(s,t)=True$. 
This case is referred to as \textit{intra-buffer checking}.
If intra-buffer checking cannot determine the query result, then we check whether $s$ and $t$ are connected via all vertices in $b_i[j]$ or $f_{i+1}[j-1]$, referred to as \textit{inter-buffer checking}.

\begin{example}\label{example:intra_and_inter_buffer_checking}
For an example of inter-buffer checking, consider query $Q_c(C,G)$ in Figures \ref{fig:bic} and \ref{fig:backward_forward_buffer}.
$Q_c(C,G)$ over $\window_3$ is evaluated using $b_1[2]$ and $f_2[1]$ according to the BIC model.
In $b_1[2]$, $C$ and $G$  are connected as they have the same root $A$, such that $Q_c(C,G)$ is $True$ in $\window_3$. 
Consider query $Q_c(I,C)$ in $\window_3$ in Figure \ref{fig:backward_forward_buffer} for an example of inter-buffer checking.
In this case, intra-buffer checking can not determine the query result because $I$ and $C$ are not connected in either $b_1[2]$ or $f_2[1]$.
Thus, inter-buffer checking is necessary. 
In $f_2[1]$, there exist vertices $D$ and $B$, which have the same root as $I$ and $C$, respectively, and $D$ and $B$ are connected in $b_1[2]$.
Thus, $Q_c(I,C)$ is $True$ in $\window_3$.
\end{example}

Intra-buffer checking is a trivial task as it only requires performing the \textit{find} operation.
The main challenge in merging is inter-buffer checking. 
We denote the vertices that appear in both $f_{i+1}[j-1]$ and $b_i[j]$ as \textit{inter-vertices}.
Then, inter-buffer checking requires searching inter-vertices transitively as they can make $s$ and $t$ connected.
For instance, in Figure \ref{fig:backward_forward_buffer}, for $b_1[2]$ and $f_2[1]$, vertices $D$ and $B$ are inter-vertices as they appear in both $b_1[2]$ and $f_2[1]$, which can make $I$ and $C$ connected, as discussed in Example \ref{example:intra_and_inter_buffer_checking}.

Searching for such inter-vertices during inter-buffer checking is computationally expensive because a UFT can have thousands or millions of vertices. More importantly, this procedure needs to be performed recursively, \textit{i.e.}, if $s$ and $t$ are not connected in $f_{i+1}[j-1]$ and the inter-vertices $s'$ and $t'$ that are respectively connected to $s$ and $t$ in $f_{i+1}[j-1]$ are still not connected in $b_i[j]$, then it is necessary to check whether there exist inter-vertices that can make $s'$ and $t'$ connected, and so forth. Such kind of checking needs to be performed exhaustively to determine the final query result. 

\subsection{Indexing for inter-buffer checking}\label{sec:BFBG}
In order to support efficient inter-buffer checking, we store the connection via inter-vertices between $b_i[j]$ and $f_{i+1}[j-1]$, $\forall j\geq 1$. 
The main idea is to store the connection between $b_i[j]$ and $f_{i+1}[j-1]$ by recording the roots of inter-vertices.
For instance, in Figure \ref{fig:backward_forward_buffer}, inter-vertex $C$ has root $A$ in $b_1[2]$ and root $K$ in $f_2[1]$. 
Thus, the connected information between $A$ and $K$ via $C$ is recorded, which will be used for inter-buffer checking.
Storing the connection between the roots of inter-vertices is more efficient than simply storing inter-vertices. The reason is that the number of inter-vertices can be large but the number of their roots in $b_i[j]$ and $f_{i+1}[j-1]$ is very small (the number of CCs is not large in practice). Consequently, a significant amount of inter-vertices can have the same root, and storing the roots will naturally remove redundancy.

We use a backward-forward bipartite graph (BFBG) to store the connection between  $b_i[j]$ and $f_{i+1}[j-1]$. 
The vertices in a BFBG are the roots in $b_i[j]$ and $f_{i+1}[j-1]$, and the edges represent the connection between the roots via inter-vertices. 
Each BFBG is computed for each pair of $(b_i, f_{i+1})$, and the BFBG is updated as the snapshot index $j$ increases. The snapshots of the BFBG can be used to provide the connectivity information between $b_i[j]$ and $f_{i+1}[j-1]$. 
In order to efficiently compute BFBGs, edges in BFBGs are labeled with additional information. 

\begin{definition}
    A BFBG $(V_b,V_f,E_{b,f})$ for $(b_i,f_{i+1})$ records the connection between roots of UFTs in $(b_i[j], f_{i+1}[j-1]),j\geq 1$.
    $V_b$ and $V_f$ are the subsets of vertices in $b_i[j]$ and $f_{i+1}[j-1]$, respectively. 
    If there exists an inter-vertex $v$ between $b_i[j]$ and $f_{i+1}[j-1]$, there must exist an edge $e_{b,f} = (v_b,v_f)\in E_{b,f}$, such that $v_b$ and $v_f$ are the roots of $v$ in $b_i[j]$ and $f_{i+1}[j-1]$, respectively.    
    In addition, each $e_{b,f}$ is labeled with one or multiple intervals $[j_s,j_e]$ such that the root of the inter-vertex $v$ is $v_b$ in $b_i[j], \forall j\in [j_s,j_e]$.     
\end{definition}

\begin{figure}[h]
    \centering
    \resizebox{0.9\linewidth}{!}{
    \includegraphics{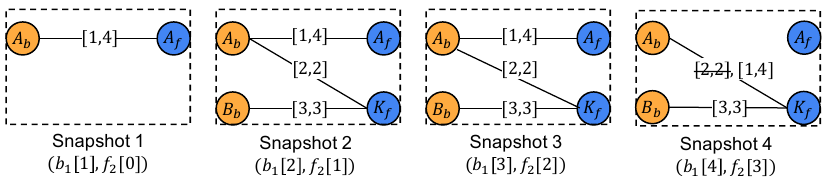}}
    \caption{The snapshots of BFBG for $b_1$ and $f_2$ in Figure \ref{fig:backward_buffer_labeling}.}
    \label{fig:BFBG}
\end{figure}

\begin{example}\label{example:illustration_BFBG}
The BFBG for $(b_1[1],f_2[0])$ of the running example is shown in snapshot $1$ in Figure \ref{fig:BFBG}. 
As the roots of inter-vertex $A$ in $b_1[1]$ and $f_2[0]$ are $A$ and $A$, respectively. Thus, edge $(A_b,A_f)$ is included in the BFBG.
In addition, according to $b_1$, the root of $A$ is $A$ from $b_1[1]$ to $b_1[4]$. Thus, $(A_b,A_f)$ in snapshot 1 of Figure \ref{fig:BFBG} is labeled with $[1,4]$.   
\end{example}

In BIC, a BFBG is created for each pair of $(b_i, f_{i+1})$, which will be used to perform inter-buffer checking between $b_i[j]$ and $f_{i+1}[j-1], j\geq 1$.
We say the $j$-th snapshot of the BFBG is the snapshot of the BFBG that can be used to perform inter-buffer checking for $(b_i[j],f_{i+1}[j-1])$, \textit{e.g.}, Figure \ref{fig:BFBG} shows the $4$ snapshots of the BFBG for $(b_1,f_2)$ in the running example.
We note that the intervals assigned to edges in a BFBG are used to deal with the issues of computing and updating the BFBG, and the corresponding detail will be explained later.
We discuss below how a BFBG can be used, followed by how to compute a BFBG.

\textit{Inter-buffer checking using a BFBG}.
The main idea of performing inter-buffer checking using a BFBG is to compute for $s$ and $t$ in $Q_c(s,t)$ their roots in $b_i[j]$ and $f_{i+1}[j-1]$, and then check whether the roots are connected in the BFBG. 
In checking connectivity using the $j$-th snapshot of the BFBG, only edges labeled with an interval such that $j$ is in the interval are visited. 

\begin{example}
Consider $Q_c(I, C)$ in $\window_3$ in Figure \ref{fig:backward_forward_buffer}.
$I$ and $C$ are not connected in $b_1[2]$ and $f_2[1]$.
Then, snapshot 2 of the BFBG for $(b_1, f_2)$ shown in Figure \ref{fig:BFBG} is used to perform inter-buffer checking.
The computation starts with finding the roots of $I$ and $C$ in $f_2[1]$, \textit{i.e.}, $A$ and $K$, respectively, and then checks whether $A_f$ and $K_f$ are connected in snapshot 2 of the BFBG.
We have $Q_c(I,C)=True$ as $A_f$ and $K_f$ are connected.
In this example, edge $(A_b, A_f)$ in snapshot 2 of the BFBG is visited because it has an interval $[1,4]$ and $2\in [1,4]$. Edge $(A_b, K_f)$ is visited because of the same reason. However, edge $(B_b, K_f)$ is pruned because $(B_b, K_f)$ is labeled with only one interval $[3,3]$ and $2\not\in [3,3]$.
\end{example}

\textit{Computing a BFBG}.
The BFBG for each pair of $(b_i, f_{i+1})$ is computed incrementally in the sense that the $j$-th snapshot is computed by performing updates on top of the $(j-1)$-th snapshot, where the first snapshot is computed on top of the empty BFBG. 
There exist two kinds of updates in a BFBG: edge insertions and updating $v_f\in V_f$. Edge insertion updates are for the case that a vertex is identified as an inter-vertex, such that the connection between $b_i[j]$ and $f_{i+1}[j-1]$ via the inter-vertex needs to be recorded. 
Updating $v_f\in V_f$ is used to deal with the issue that the root of an inter-vertex in $f_{i+1}[j-1]$ is changed as $j$ increases (this is because of processing new streaming edges).
One may notice that the root of an inter-vertex in $b_i[j]$ may also be changed as $j$ increases. 
However, such kind of updates do not need to be performed because of the interval assignment mechanism on the edges in BFBGs. 
The intuition is to insert all the possible roots of an inter-vertex $v$ in different snapshots of $b_i$ when $v$ is identified as an inter-vertex and to distinguish the roots using the assigned intervals.
This tailored design is to avoid recomputing the roots of inter-vertices in $b_i[j]$ for different $j$. 
We note that it is feasible to have all the roots of an inter-vertex in $b_i[j], \forall j \geq 1$ because $b_i$ is computed in the backward manner.

\textit{Edge insertions in a BFBG}. 
For each streaming edge inserted in $f_{i+1}[j-1]$, we check whether each of the two endpoints of the streaming edge is an inter-vertex.
If $v$ is identified as an inter-vertex with root $v_f$ in $f_{i+1}[j-1]$ and root $v_b$ in $b_i[j]$, edge $(v_b,v_f)$ is inserted into the current snapshot of the BFBG, which is labeled with $[j_s,j_e], j_s=j$, such that $v_b$ is the root of $v$ from $b_i[j_s]$ to $b_i[j_e]$.
If inter-vertex $v$ has different roots in $b_i[j'], j'>j_e$, for each such root $v'_b$, edge $(v'_b,v_f)$ labeled with $[j'_s,j'_e]$ is inserted, such that $v'_b$ is the root of $v$ from $b_i[j'_s]$ to $b_i[j'_e]$.
These additional edge insertions allow the BFBG to keep track of changes of the roots of inter-vertices in $b_i$ and avoid recomputing the roots.
We note that computing all possible intervals and the corresponding roots in $b_i$ is feasible (details are discussed in \S \ref{sec:aug_backward_uft}). 

\begin{example}\label{example:illustration_BFBG_multiple_insertion}
Snapshot $2$ of the BFBG in Figure \ref{fig:BFBG} is computed by inserting edges into snapshot $1$.
$C$ is identified as an inter-vertex, which has root $A$ in $b_1[2]$ and root $K$ in $f_2[1]$, such that edge $(A_b,K_f)$ is inserted.
In addition, $C$ has root $B$ in $b_1[3]$. 
Thus, $(A_b,K_f)$ is labeled with $[2,2]$, and another edge $(B_b,K_f)$ labeled with $[3,3]$ is also inserted.
Notice that $C$ does not exist in $b_1[4]$ and thus it is not an inter-vertex in $(b_1[4], f_2[3])$. 
Such kind of information is recorded in BFBG by the corresponding inserted intervals, \textit{i.e.}, $4$ is neither in $[2,2]$ nor in $[3,3]$.
Inter-vertex $B$ has the same roots and the same intervals as inter-vertex $C$, such that no edge is inserted into the BFBG for $B$.
\end{example}

When inserting edge $(v_b,v_f)$ with interval $[j_s,j_e]$, $(v_b,v_f)$ might already exist in BFBG but labeled with a different interval $[j'_s,j'_e]$.
In this case, if  $[j_s,j_e]$ and $[j'_s,j'_e]$  overlap, they can be merged to condense BFBG, \textit{e.g.}, for edge $(A_b,K_f)$ in snapshot 4 in Figure \ref{fig:BFBG}, $[2,2]$ is subsumed by $[1,4]$ such that $[2,2]$ can be deleted.

\textit{Updating $v_f$ in edge $(v_b,v_f)$ in a BFBG}.
When a new streaming edge $e$ is inserted into $f_{i+1}$, the \textit{union} operation will link the roots of the two endpoints of $e$ in $f_{i+1}$.
Assuming root $v$ is linked as a child root of $u$. In this case, if $v$ also exists in $V_f$ in the current snapshot of the BFBG, the edges adjacent to $v$ need to be moved to $u$.
Consider $f_2[2]$ and $f_2[3]$ in Figure \ref{fig:backward_forward_buffer}. Both $A$ and $K$ are roots in $f_2[2]$, but $A$ becomes a child of $K$ in $f_2[3]$. Thus, in Figure \ref{fig:BFBG}, edge $(A_b, A_f)$ in snapshot 3 is changed to edge $(A_b, K_f)$ in snapshot $4$.

\subsection{Augmented UFTs in backward buffers}\label{sec:aug_backward_uft}
We discuss how to compute the intervals when inserting edges into a BFBG. 
Interval $[j_s,j_e]$ of edge $(v_b,v_f)$ indicates that $v_b$ is the root of an inter-vertex $v$ from $b_i[j_s]$ to $b_i[j_e]$. 
Obviously, it is computationally expensive to have the intervals by computing the root of $v$ in each snapshot of $b_i$, because the \textit{find} operation in the UFT has to be called repeatedly, \textit{i.e.}, $O(|c|)$ times. 
We aim at performing the \textit{find} operation once to get all the possible roots and the corresponding intervals. 
This is possible because backward buffer $b_i$ is computed incrementally and in the backward manner.
Therefore, all the information related to the changes of the roots in $b_i$ can be obtained.
To achieve this goal, we augment UFTs in each $b_i$ with additional information, and such UFTs are denoted as augmented UFTs (AUFTs), defined below.

\begin{definition}
Given a backward buffer $b$, augmented UFTs (AUFTs) in each snapshot in $b$ records the following information:
\begin{itemize}
    \item Each vertex $v$ in an AUFT is labeled with a snapshot index $j$ in $b$, such that $j$ is the largest of all the snapshot indexes of $b$, which contains $v$, \textit{i.e.}, $j=max(\{j'|b[j'] \text{ contains } v\})$.
    \item Each root $r$ in an AUFT is labeled with an interval $[1,j_e]$, such that $j_e$ is the largest of all the snapshot indexes of $b$, where $r$ is a root,  \textit{i.e.}, $j_e=max(\{j'|r \text{ is a root in } b[j']\})$. 
    \item  If vertex $v$ is a root labeled with $[1,j_e]$ in $b[j]$ but $v$ is not a root in $b[j-1]$,  then $v$ is labeled with interval $[j,j_e]$.
\end{itemize}
\end{definition}

\begin{figure}[h]
    \centering
    \resizebox{0.9\linewidth}{!}{
    \includegraphics{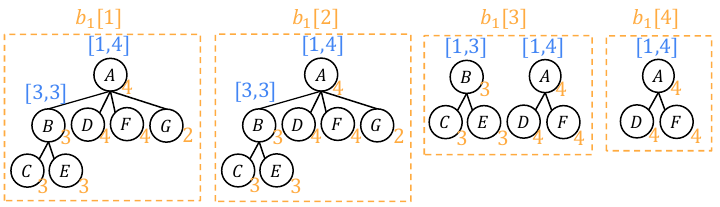}}
    \caption{The AUFTs in backward buffer $b_1$ in Figure \ref{fig:backward_forward_buffer}.}
    \label{fig:agumented_backward_trees}
\end{figure}

\begin{example}\label{example:AUFTs}
We present in Figure \ref{fig:agumented_backward_trees} the AUFTs in the snapshots of $b_1$ in the running example of Figure 3.
Vertices $A$, $D$, and $F$ are labeled with $4$ because $4$ is the largest snapshot index.
Vertex $B$ is labeled with interval $[1,3]$ in $b_1[3]$ as $3$ is the largest snapshot index such that $B$ is a root.
In $b_1[2]$, the interval of $B$ is changed to $[3,3]$ because $B$ becomes a child of $A$.
\end{example}

Although AUFTs record additional information compared to UFTs, AUFTs can be computed and stored in the same way as UFTs.
The intuition is that backward buffer $b$ is computed in the backward manner, such that the vertex label of $v$ corresponds to the first snapshot where $v$ appears, and $j_e$ in interval $[1,j_e]$ of root $r$ corresponds to the first snapshot where $r$ is a root.
For instance, in Example \ref{example:AUFTs}, $b_1[4]$ is first computed, where $A$ appears and $A$ is also a root, such that $A$ has a vertex label $4$ and interval $[1, 4]$.

We note that snapshots of backward buffer $b$ with AUFTs can be stored using the snapshot isolation approach presented in \S \ref{sec:backward_storage} because vertex labels and root intervals can be naturally stored with vertices and roots in AUFTs.

We discuss below how to use the augmented information in AUFTs to compute the roots of inter-vertex $v$ in the snapshots of $b_i$. 
The main idea is to retrieve the path from $v$ to the root of $v$ in the current snapshot $b_i[j]$ and then leverage the vertex label of $v$ and root intervals of vertices along the path to compute the roots of $v$ in all the snapshots $b_i[j'], j'\geq j$.
This is possible because the vertex label of $v$ indicates when $v$ exists in the snapshots of $b_i$ and the root interval of any vertex $u$ along the path indicates when $u$ is a root in the snapshots of $b_i$.  Thus, we can derive the root of $v$ at each snapshot of $b_i$.
For ease of presentation, we use the running example to explain this procedure below.

\begin{example}
  Consider Example \ref{example:illustration_BFBG_multiple_insertion}, where $C$ is identified as an inter-vertex between $b_1[2]$ and $f_2[1]$, and the root of $C$ in $f_2[1]$ is $K$.  
  In order to insert edges into the BFBG for $(b_1[2],f_2[1])$, the AUFT stored in $b_1[2]$ shown in Figure \ref{fig:agumented_backward_trees} is used.
  The path from $C$ to the root in the AUFT is $(C,B,A)$, where $B$ and $A$ have intervals in the AUFT. 
  The label of $C$ is $3$, indicating that $C$ is inserted at $b_1[3]$, and the interval of $B$ is $[3,3]$, indicating that $B$ is a root in $b_1[3]$. Therefore, we can derive that $B$ is the root of $C$ in $b_1[3]$, which leads to inserting bipartite edge $(B_b, K_f)$ with $[3,3]$ in snapshot $2$ of the BFBG in Figure \ref{fig:BFBG}.
  The next vertex in the path $(C,B,A)$ is $A$ labeled with $[1,4]$, indicating that $A$ is a root from $b_1[1]$ to $b_1[4]$.
  Since the current snapshot index is $2$ (when $C$ is identified as an inter-vertex) and $C$ has root $B$ in $b_1[3]$, such that we can derive that $A$ is the root of $C$ in $b_1[2]$, which leads to inserting bipartite edge $(A_b, K_f)$ with $[2,2]$ in snapshot $2$ of the BFBG in Figure \ref{fig:BFBG}.
\end{example}

\subsection{End-to-end computation \&\ complexity}\label{sec:backward_parallel}
Consider query $Q_c(s,t)$ over sliding windows.
Let $n$ and $m$ be the number of vertices and edges in a window, respectively. 

\textit{Query time}.
A simple case in query processing is a forward buffer stores the connectivity information of the entire window, \textit{e.g.}, $f_2[4]$ for $\window_6$ in Figure \ref{fig:bic}.
In this case, $Q_c(s,t)$ can be simply processed by checking whether $\rt{s}$ and $\rt{t}$ are the same in the forward buffer, which takes at most $O(\log{n})$ time (Lemma \ref{lemma:optimized_uft}).
In other cases, the merging operation is necessary.
Intra-buffer checking in $f_{i+1}$ or $b_i$ is first applied, taking at most $O(\log{n})$ time.
If the query result cannot be determined, inter-buffer checking is performed by using the current snapshot of the BFBG for $b_i$ and $f_{i+1}$, taking $O((|V_b|+|V_f|+|E_{b,f}|)\log{|c|})$ time (the intervals of each bipartite edge can be stored using a interval tree \cite{wiki:Interval_tree} and each edge has at most $\log{|c|}$ intervals). 
Thus, the total query time is $O(\log{n} + C)$ in the worst case, where $C$ is $(|V_b|+|V_f|+|E_{b,f}|)\log{|c|}$.
Notice that, $|V_b|$ and $|V_f|$ are at most the number of CCs in $b_i$ and $f_{i+1}$, respectively, which are very small in practice. Thus, $|E_{b,f}|$, bounded by $|V_b|\times|V_f|$, is also very small. 
$|c|$ is window size divided by slide interval, which is small in practical settings \cite{10.1145/3318464.3389733}.
Consequently, $C$ can be negligible, and the query time can be $O(\log{n})$ in practice.

\textit{Update time}.
After receiving a streaming edge $e=(u,v)$, we need to update the forward buffer, the backward buffer, and the BFBG between them.
The update in forward buffer $f$ inserts $e$ into the maintained UFTs, which takes $O(\log n)$ time.
If the insertion requires performing the \textit{union} operation, we might also need to reflect the changes in the current BFBG (see updating $v_f$ in $(v_b,v_f)$ in \S \ref{sec:BFBG}). 
The corresponding change takes at most $O(|V_B||c|\log{|c|})$ time because there can be at most $|V_b|$ edges adjacent to a vertex in $V_f$, each edge can have at most $|c|$ intervals, and inserting each interval takes $O(\log{|c|})$ time.
If the current chunk is full, the backward buffer needs to be computed over the chunk, taking $O(m\log{n})$ time because of scanning all the edges in the chunk. 
If the chunk is not full and $u$ (or $v$) is an inter-vertex between $b_i[j]$ and $f_{i+1}[j-1]$, the updates in the current BFBG are needed.
It takes at most $O(\log{|n|})$ time to compute the roots of inter-vertices in $b_i$, and there can be at most  $O(\log{|n|})$ roots. Each root corresponds to an edge insertion into the BFBG, taking at most $O(\log{|V_b|} + \log{|V_f|} + \log{|c|})$ time.
Thus, the update for processing $e$ takes at most
$O(C' + m\log{n})$ or $O(C' + C''\log{n})$ time, where $C'=|V_B||c|\log{|c|}$ and $C''=\log{|V_b|} + \log{|V_f|} + \log{|c|}$. 
As $|V_f|$, $|V_b|$, and $|c|$ are very small compared to $n$ and $m$ in practice, the update can take $O(m\log{n})$ or $O(\log{n})$ time.

Notice that the computation of the backward buffer is performed per chunk, such that the cost of $O(m\log{n})$ is amortized over all edges in a chunk. Thus, the amortized update time is $O(\log{n})$. 
It is also noteworthy that our algorithm does not need to delete expired  edges from each window.  
Deleting an expired edge takes $O(n+m)$ time in the worst case using the FDC approach (see \S \ref{sec:related_works}), and each slide interval can have millions of edges, all of which need to be deleted.
To sum up, the near $O(\log{n})$ worst-case query time and amortized update time demonstrate the high-throughput and low-latency properties of our approach, which is a fundamental requirement in stream processing.

\textit{Space}.
BIC's space complexity is $O(|c|n+m)$. 
Each backward or forward buffer necessitates at most $O(n)$ space, with BFBG containing at most $O(n)$ edges, each spanning up to $|c|$ intervals. Additionally, storing edges in each chunk is essential for computing the backward buffer of the chunk.
The space complexity of BIC mirrors that of FDC approaches \cite{10.1145/320211.320215,10.1145/502090.502095,10.14778/3551793.3551868}, which also require $O(n + m)$ space. 
However, FDC methods store all edges per window instance, while BIC stores all edges per chunk, leading to BIC's lower memory usage, as demonstrated in our experiments (\S \ref{sec:memory_udage}).

\section{experimental evaluation}\label{sec:experiments}
We evaluate  BIC against current state-of-the-art methods using $8$ real-world datasets and $2$ synthetic datasets from industrial-grade benchmarks. 
Our evaluation includes throughput and latency analysis in \S \ref{sec:performance}, emphasizing windows of a few million edges. 
We then examine the settings with various window sizes and slide intervals in \S \ref{sec:scalability}, focusing on large windows of up to $80$ million edges. 
We analyze the impact of workload size in \S \ref{sec:workload}.
Finally, we analyze the memory usage in all these settings.

\begin{figure*}
\begin{minipage}{0.095\linewidth}
\resizebox{\textwidth}{!}{
    \includegraphics{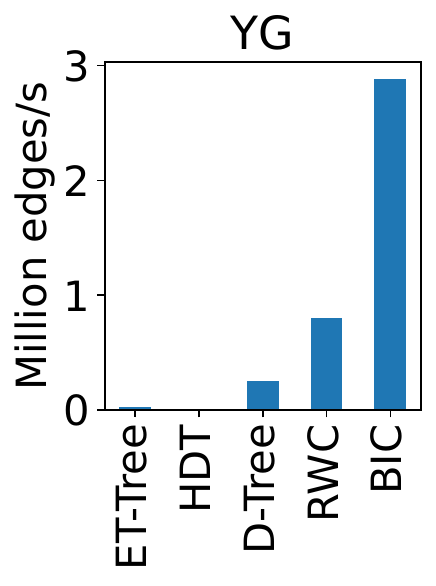}
    }    
\end{minipage}
\begin{minipage}{0.095\linewidth}
\resizebox{\textwidth}{!}{
    \includegraphics{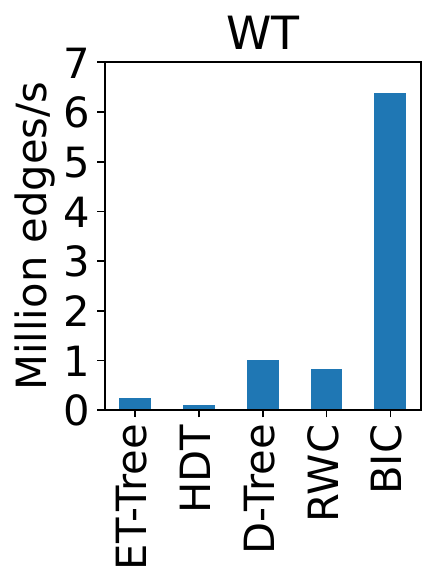}
    }
\end{minipage}
\begin{minipage}{0.095\linewidth}
\resizebox{\textwidth}{!}{
    \includegraphics{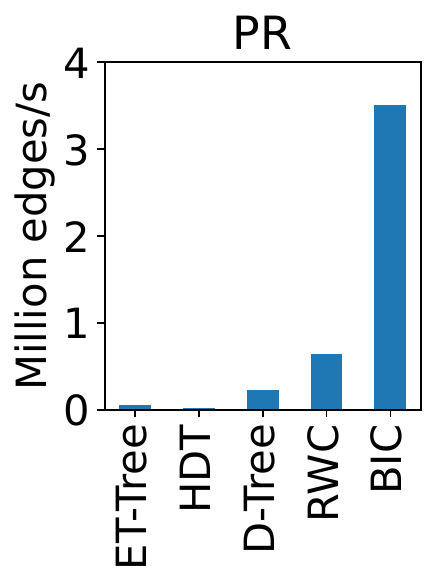}
    }
\end{minipage}
\begin{minipage}{0.095\linewidth}
\resizebox{\textwidth}{!}{
    \includegraphics{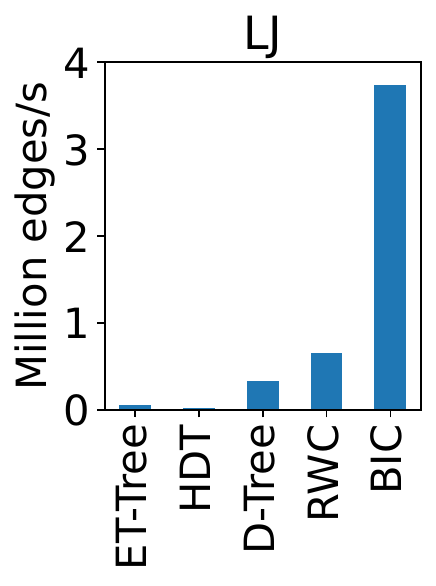}
    }
\end{minipage}
\begin{minipage}{0.095\linewidth}
\resizebox{\textwidth}{!}{
    \includegraphics{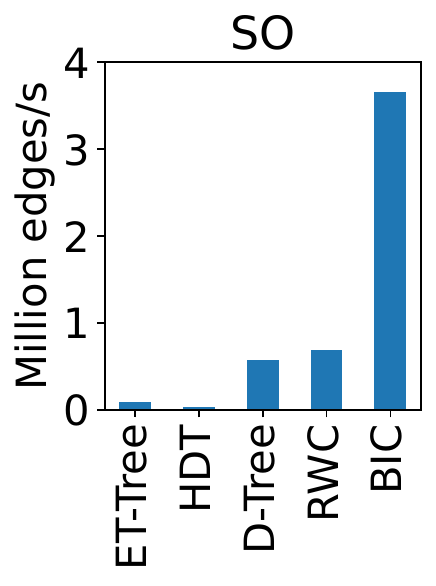}
    }
\end{minipage}
\begin{minipage}{0.095\linewidth}
\resizebox{\textwidth}{!}{
    \includegraphics{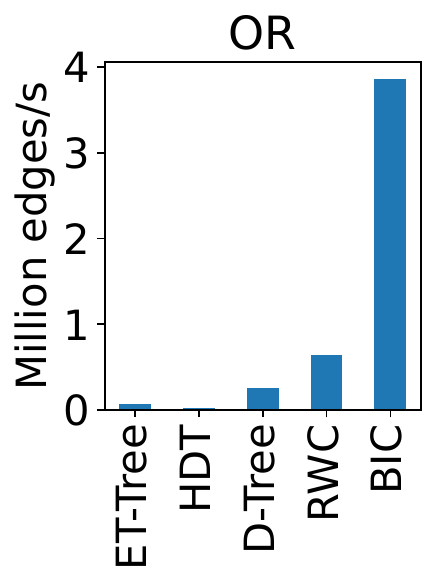}
    }
\end{minipage}
\begin{minipage}{0.095\linewidth}
\resizebox{\textwidth}{!}{
    \includegraphics{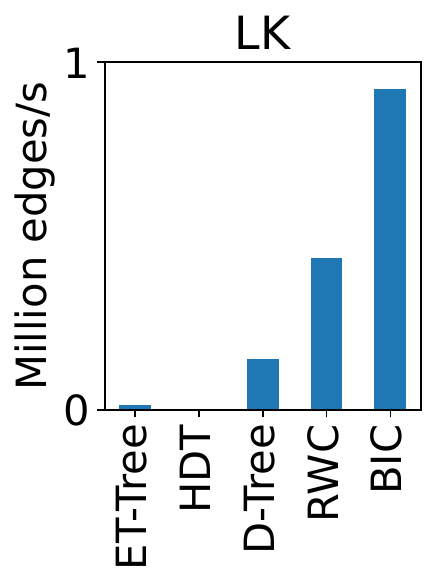}
    }
\end{minipage}
\begin{minipage}{0.095\linewidth}
\resizebox{\textwidth}{!}{
    \includegraphics{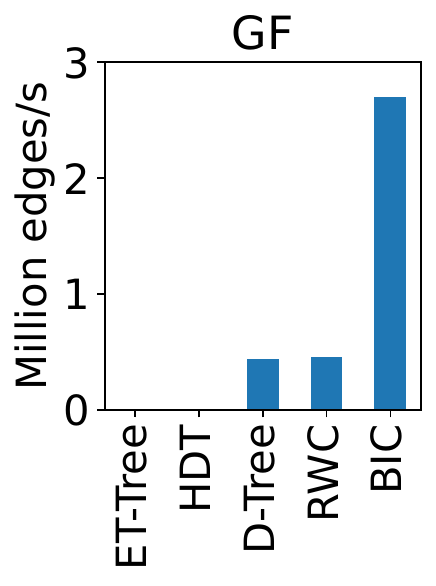}
    }
\end{minipage}
\begin{minipage}{0.095\linewidth}
\resizebox{\textwidth}{!}{
    \includegraphics{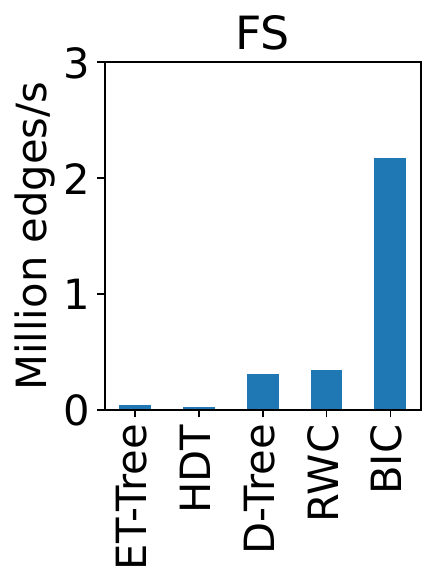}
    }
\end{minipage}
\begin{minipage}{0.095\linewidth}
\resizebox{\textwidth}{!}{
    \includegraphics{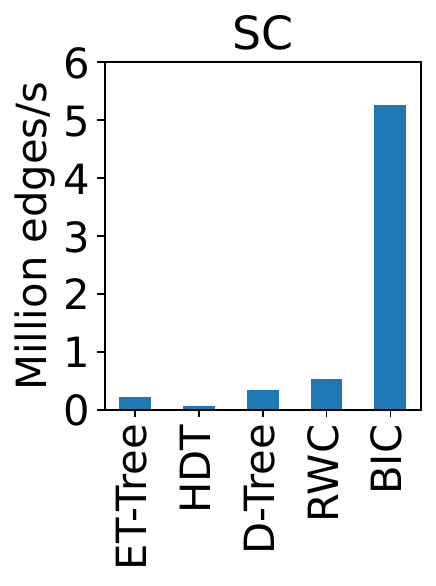}
    }
\end{minipage}
\caption{Throughput analysis using windows of on average $3$M edges and slide intervals of on average $150$K edges.}\label{fig:per_throughput}
\end{figure*}

\begin{figure*}
\centering
\resizebox{0.25\textwidth}{!}{
\includegraphics{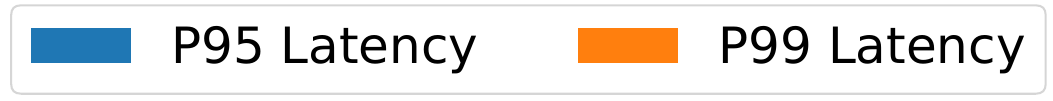}
} \\
\begin{minipage}{0.095\linewidth}
\resizebox{\textwidth}{!}{
    \includegraphics{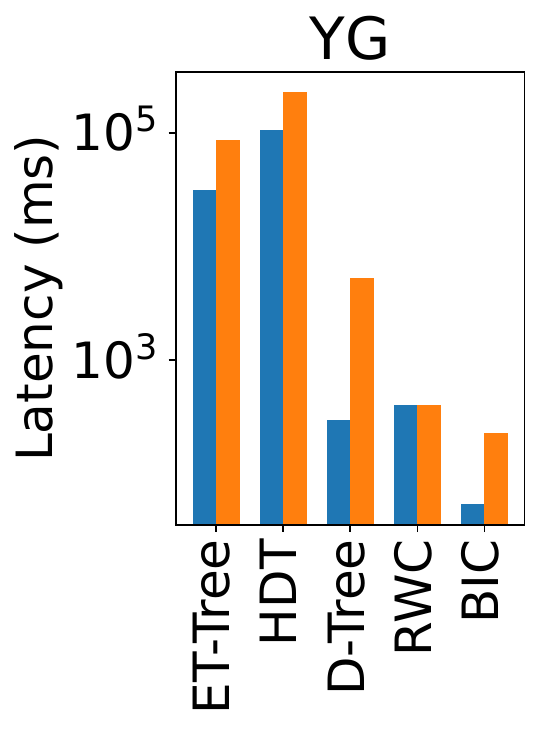}
    }
\end{minipage}
\begin{minipage}{0.095\linewidth}
\resizebox{\textwidth}{!}{
    \includegraphics{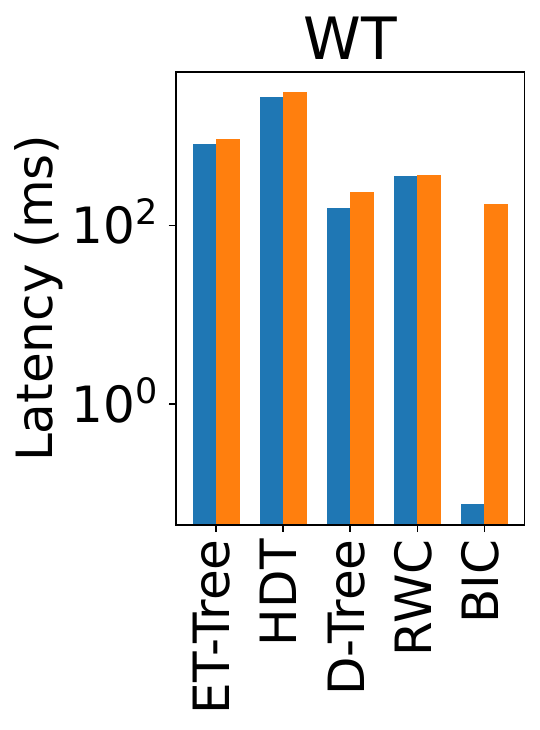}
    }
\end{minipage}
\begin{minipage}{0.095\linewidth}
\resizebox{\textwidth}{!}{
    \includegraphics{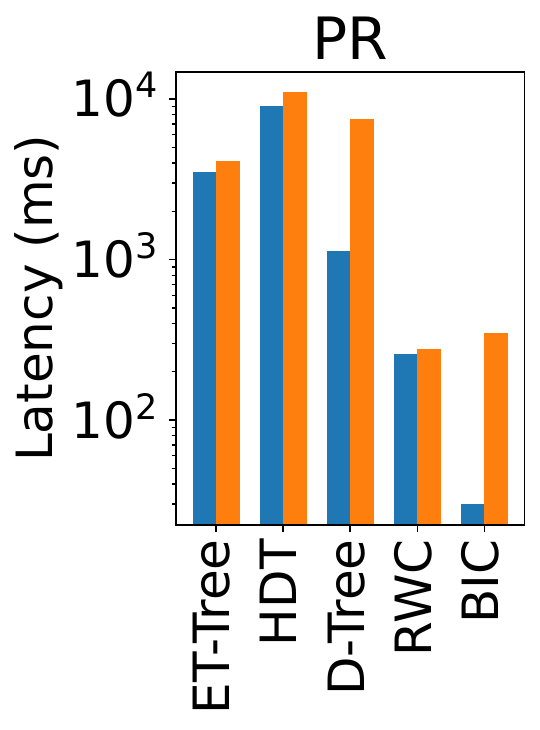}
    }
\end{minipage}
\begin{minipage}{0.095\linewidth}
\resizebox{\textwidth}{!}{
    \includegraphics{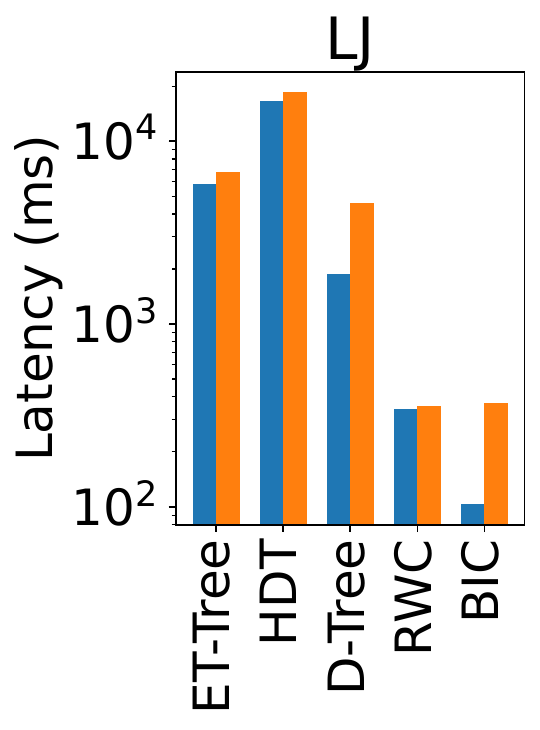}
    }
\end{minipage}
\begin{minipage}{0.095\linewidth}
\resizebox{\textwidth}{!}{
    \includegraphics{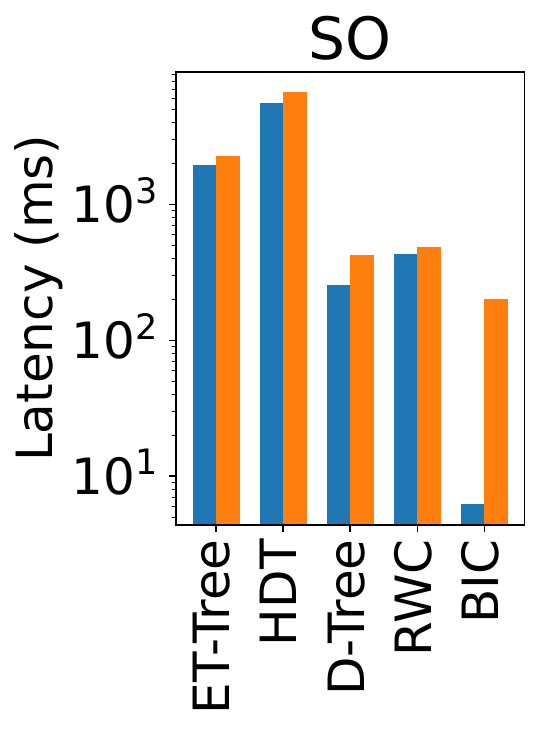}
    }
\end{minipage}
\begin{minipage}{0.095\linewidth}
\resizebox{\textwidth}{!}{
    \includegraphics{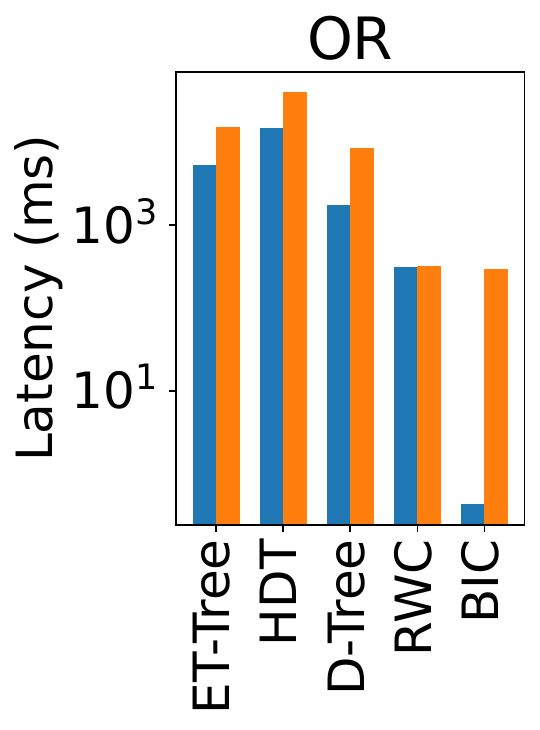}
    }
\end{minipage}
\begin{minipage}{0.095\linewidth}
\resizebox{\textwidth}{!}{
    \includegraphics{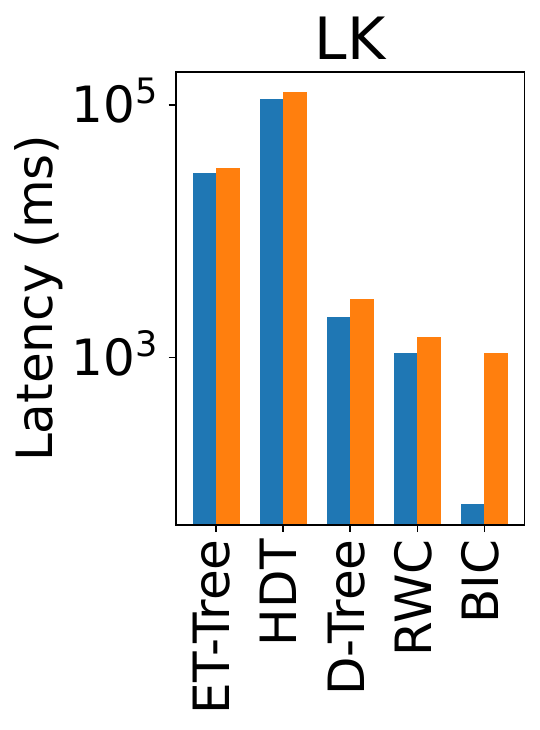}
    }
\end{minipage}
\begin{minipage}{0.095\linewidth}
\resizebox{\textwidth}{!}{
    \includegraphics{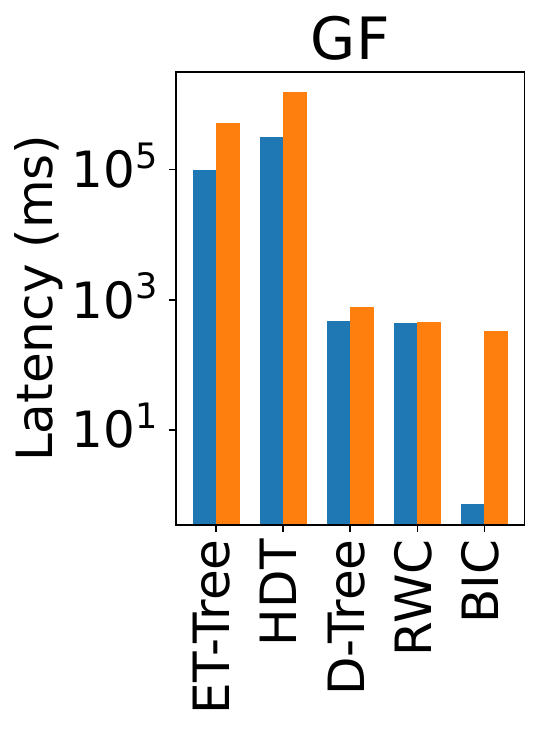}
    }
\end{minipage}
\begin{minipage}{0.095\linewidth}
\resizebox{\textwidth}{!}{
    \includegraphics{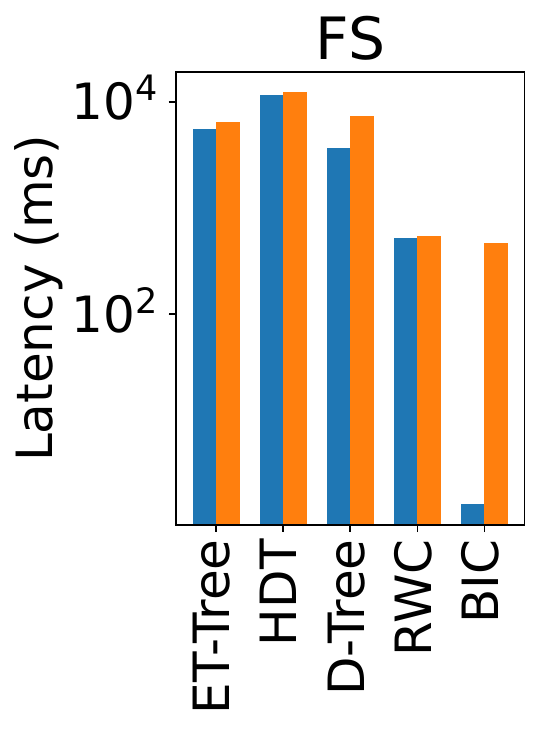}
    }
\end{minipage}
\begin{minipage}{0.095\linewidth}
\resizebox{\textwidth}{!}{
    \includegraphics{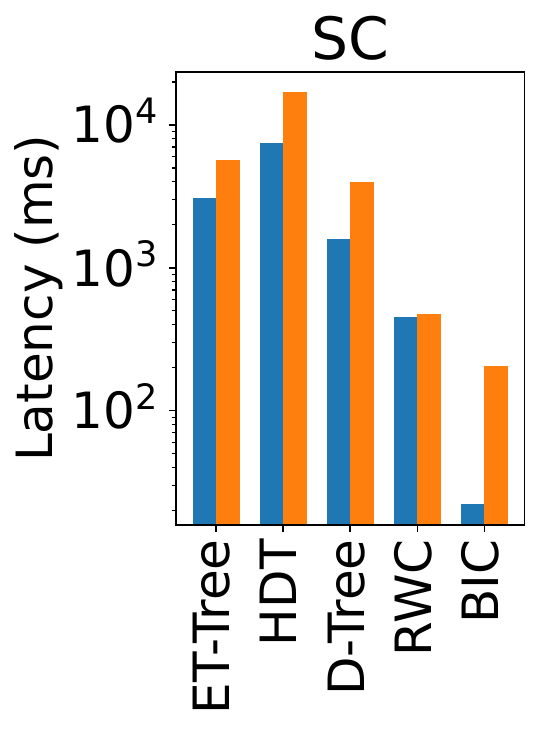}
    }
\end{minipage}
\caption{P95 and P99 tail latency analysis using windows of on average $3$M edges and slide intervals of on average $150$K edges.}\label{fig:per_latency}
\end{figure*}

\subsection{Experimental setup}\label{sec:exp_setup}
In the experiments, we denote our approach as \textbf{BIC} and compare it against the following approaches : \textbf{Euler-Tour Tree (ET-Tree)} \cite{10.1137/0214061}, \textbf{HDT} \cite{10.1145/276698.276715,10.1145/502090.502095}, \textbf{D-Tree} \cite{10.14778/3551793.3551868},  \textbf{Depth-First Search (DFS)}, and  \textbf{Recalculating Window Connectivity (RWC)}.
ET-Tree, HDT, and D-Tree are FDC data structures, designed based on spanning trees (see \S \ref{sec:related_works} for details), and D-Tree is the current state-of-the-art approach.
DFS corresponds to executing a depth-first search for each query in each window instance.
RWC recalculates all CCs in each window instance for query processing.

\textit{Datasets and workloads}.
In the experiments, we use $10$ datasets shown in Table \ref{table:datasets}, including $8$ real-world graphs \cite{snapnets},  YG, WT, PR, LJ, SO, OR, FS, and SC,  and $2$ synthetic graphs from industrial-grade benchmarks, LK \cite{10.14778/3574245.3574270} with scale factor $1000$ and GF \cite{murphy2010introducing} with scale factor $25$.
We present average distance (AD) and diameter (D) in Table \ref{table:datasets} for those such that their AD and D are available online.
We simulate a streaming graph using each dataset. 
Edges in SO and LK have timestamps, which we use in the experiments. 
For the other datasets, we assign timestamps to edges - each timestamp is assigned to $100$ edges on average.
We randomly generate workloads of queries for each graph.
The source code on the dataset and workload setups are provided in our codebase.

\begin{table}[t]
    \centering
    \caption{Overview of datasets.}
    \label{table:datasets}
    \resizebox{0.46\linewidth}{!}{
    \begin{tabular}{|c|r|r|r|r|}
    \hline
         \textbf{Dataset}   & $|V|$ & $|E|$ &AD &D \\ \hline
         \href{http://konect.cc/networks/youtube-u-growth/}{Youtube-growth (YG)}& 3.2M & 14.4M & $5.2$ & $31$\\ \hline
          \href{https://snap.stanford.edu/data/wiki-topcats.html}{Wiki-top (WT)}     & 1.7M  & 28.5M & - & 9 \\ \hline           
          \href{http://konect.cc/networks/soc-pokec-relationships/}{Pokec (PR)}        & 1.6M  & 30.6M &4.6& 14\\ \hline
          \href{https://snap.stanford.edu/data/com-LiveJournal.html}{LiveJournal (LJ)}  & 3.9M  & 34.6M &-& 17\\ \hline
          \href{http://konect.cc/networks/sx-stackoverflow/}{StackOverflow (SO)}& 2.6M  & 63.4M &3.9 & 11 \\ \hline
        \end{tabular}}
        \resizebox{0.5\linewidth}{!}{
        
        \begin{tabular}{|c|r|r|r|r|}
        \hline
             \textbf{Dataset}   & $|V|$ & $|E|$ &AD&D\\ \hline        
         \href{http://konect.cc/networks/orkut-links/}{Orkut (OR)}           & 3M    & 117.1M&4.2&10 \\ \hline          
         \href{https://github.com/ldbc/ldbc_snb_bi/blob/main/snb-bi-pre-generated-data-sets.md}{LDBC SNB Knows (LK)}   & 3.3M  & 187.2M & - & -\\ \hline
          \href{https://ldbcouncil.org/benchmarks/graphalytics/}{Graph-500  (GF)}       & 17M   & 523.6M &-&-\\ \hline          
          \href{https://snap.stanford.edu/data/com-Friendster.html}{Friendster (FS)}   & 63.6M & 1.8B &-& 32  \\ \hline
          \href{https://www.semanticscholar.org/product/api}{Semantic Scholar (SC)} & 65M & 8.27B & - &- \\ \hline
    \end{tabular}}
\end{table}

\textit{Evaluation metrics}.
We report both throughput and tail latency in our experiments, which are crucial in stream processing systems. 
Throughput is computed as the processing time of a dataset divided by the number of edges in the dataset. 
For latency, we record the response time of each approach when a streaming edge indicates the current window is complete. This includes processing queries in the window and window updates, \textit{i.e.}, a round of \texttt{query}, \texttt{insert} and \texttt{delete}.
Due to space considerations, we do not show breakdown figures, but we can report that per-edge processing latency is the dominant overhead.
Notice that, the response time includes the execution time of the most expensive operation in each compared approach, \textit{e.g.}, computing backward buffers in BIC, computing connected components in RWC, performing traversals in DFS, and deleting expired edges in FDC-based approaches (ET-Tree, HDT, and D-Tree).
We report 99th and 95th percentile latency, referred to as P99 latency and P95 latency, respectively. These are widely used metrics to evaluate system response time.

\textit{Settings}. As queries and window updates (edge insertions and deletions) are performed in sliding windows, such that the number of edges in windows and slide intervals has an impact on the performances (throughput and tail latency) of all compared approaches.
Thus, we consider various settings with respect to the number of edges in windows and slide intervals. 
We start the experiments in \S \ref{sec:performance} with a window size and a slide interval such that each window and each slide interval contain on average $3$M edges and $150$K edges, respectively, across all datasets.
We report the throughput and tail latency of each approach for each dataset.
Then, in \S \ref{sec:scalability}, we focus on the cases where  windows increase to up to $80$M edges and  slide intervals increase up to $8$M edges. 
Specifically, we consider the following two scenarios in \S \ref{sec:scalability}.
Scenario 1  fixes the slide interval but varies the window sizes. Each slide interval contains on average $1$M edges, and windows contain on average $10$M, $20$M, $40$M, and $80$M edges, respectively. 
Scenario 2 fixes the window size but varies slide intervals. Each window contains on average $80$M edges, and slide intervals contain on average $1$M, $2$M, $4$M, and $8$M edges, respectively.    
We report throughput and tail latency (P95 latency and P99 latency) in Scenario 1 and in Scenario 2, respectively. 
We use GF and FS in the experiments in \S \ref{sec:scalability} as they have a large number of edges that allow to test different settings. 
In the experiments in \S \ref{sec:performance} and \S \ref{sec:scalability}, we focus on computing a workload of $100$ queries, and all approaches are compared except DFS as performing DFS for each query in each window instance has very poor performance.
Then, in the experiments in \S \ref{sec:workload}, we study the impact of the number of queries in workloads, where we consider $1$, $10$, $100$, $1000$, and $10000$ queries, and DFS is included.
Each window and slide interval contains on average $20$M and $1$M edges, respectively, for the experiment in \S \ref{sec:workload}.
We use these settings of window sizes and slide sizes to study the memory usage of all approaches in \S \ref{sec:memory_udage}.

The implementation of BIC, all compared approaches, and experiment setups are included in our codebase that has been made publicly available. We run the experiments on a server with Ubuntu 22.04, $80$ CPUs of $2.30$GHz, and $1$TB main memory.
We note that all approaches are single-threaded.

% \footnote{Open Source Link will be provided in camera-ready version.}

\begin{figure*}
\resizebox{0.25\linewidth}{!}{
\includegraphics{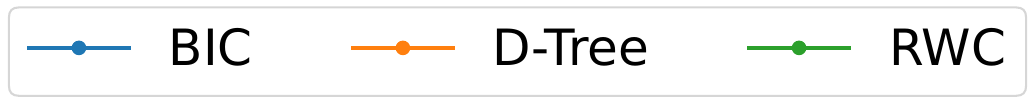}}\\
\begin{minipage}{0.29\linewidth}
    \begin{minipage}{0.32\textwidth}
        \resizebox{\textwidth}{!}{
            \includegraphics{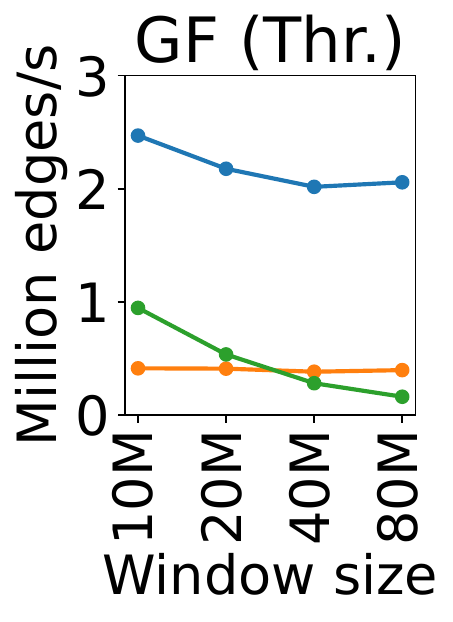}
        }
    \end{minipage}
    \begin{minipage}{0.32\textwidth}
        \resizebox{\textwidth}{!}{
            \includegraphics{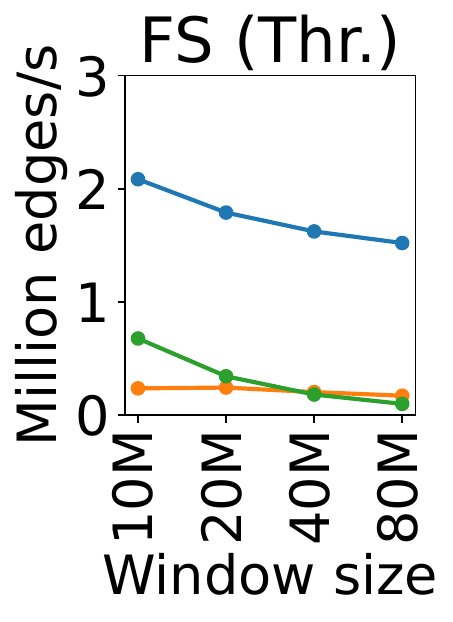}
        }
    \end{minipage}
    \begin{minipage}{0.32\textwidth}
        \resizebox{\textwidth}{!}{
            \includegraphics{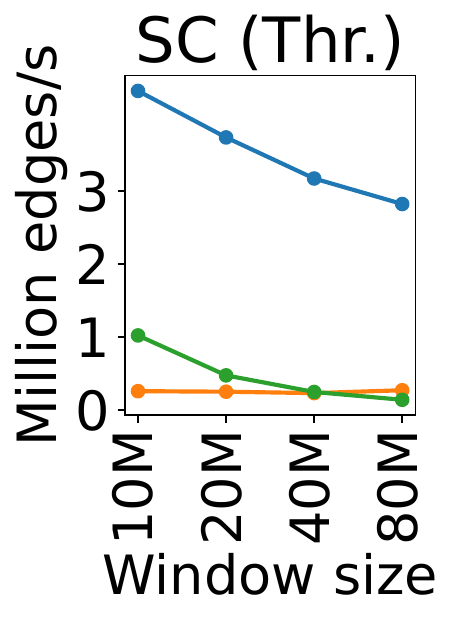}
        }
    \end{minipage}
\end{minipage}
\begin{minipage}{0.345\linewidth}
    \begin{minipage}{0.32\textwidth}
        \resizebox{\textwidth}{!}{
\includegraphics{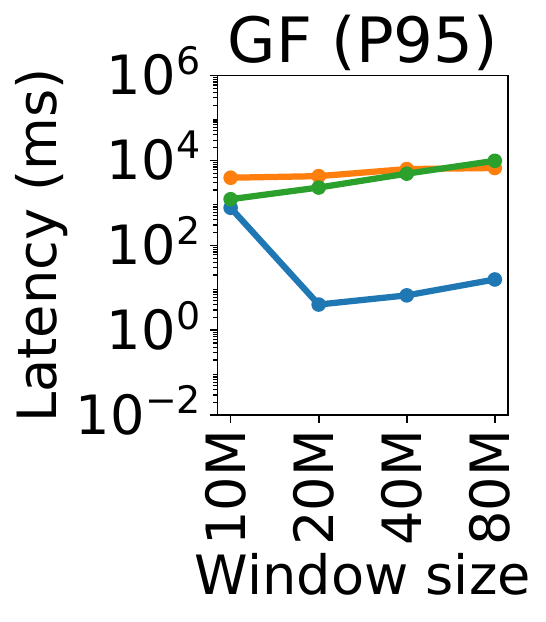}
}
    \end{minipage}
    \begin{minipage}{0.32\textwidth}
        \resizebox{\textwidth}{!}{
\includegraphics{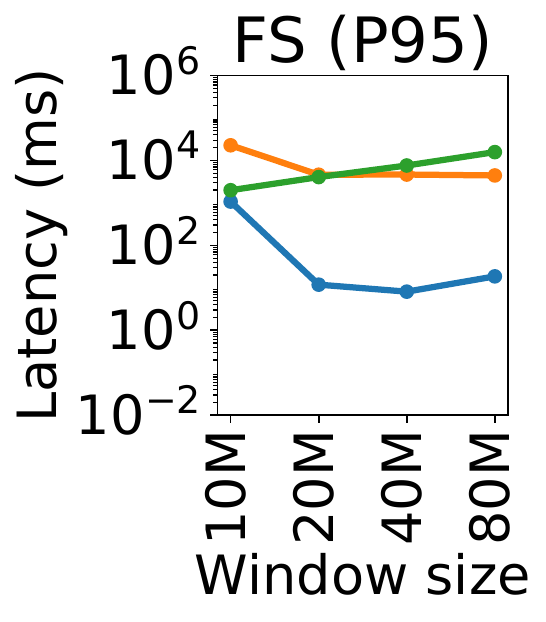}
}
    \end{minipage}
    \begin{minipage}{0.32\textwidth}
        \resizebox{\textwidth}{!}{
\includegraphics{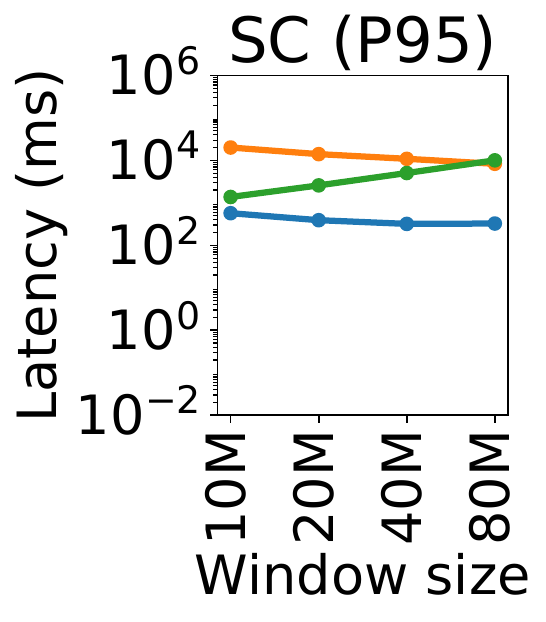}
}
    \end{minipage}
\end{minipage}
\begin{minipage}{0.345\linewidth}
    \begin{minipage}{0.32\textwidth}
        \resizebox{\textwidth}{!}{
\includegraphics{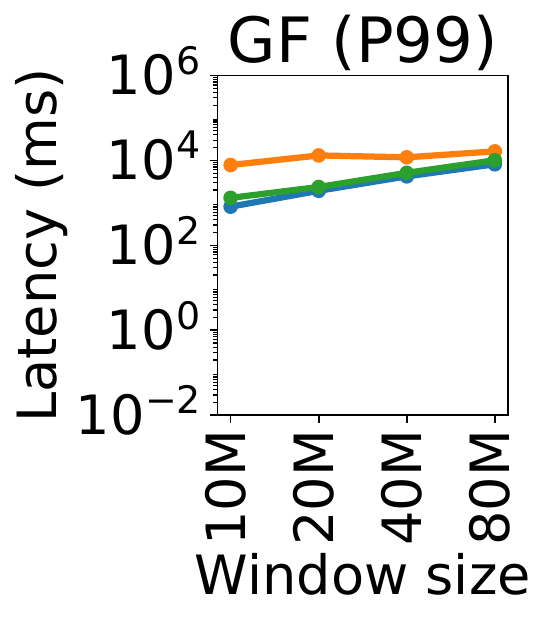}
}
    \end{minipage}
    \begin{minipage}{0.32\textwidth}
        \resizebox{\textwidth}{!}{
\includegraphics{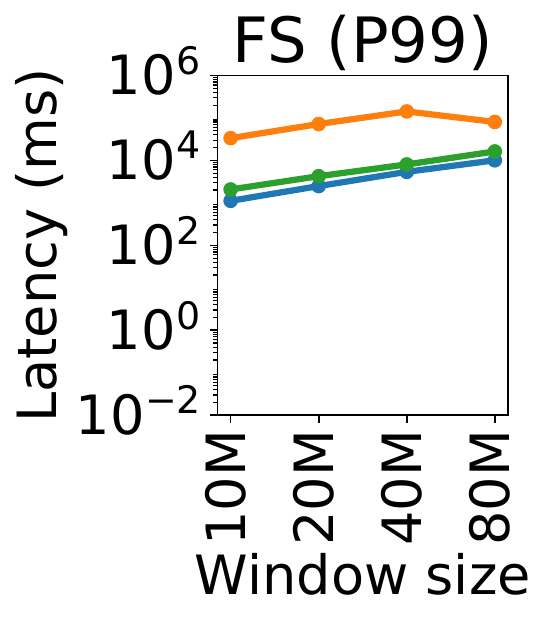}
}
    \end{minipage}
        \begin{minipage}{0.32\textwidth}
        \resizebox{\textwidth}{!}{
\includegraphics{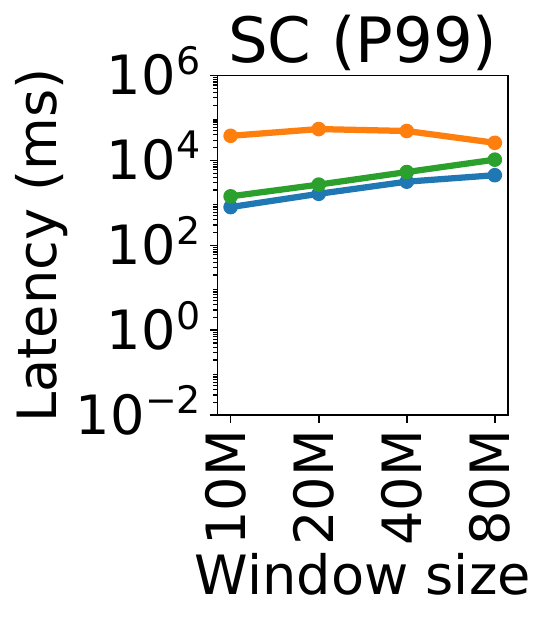}
}
    \end{minipage}
    
\end{minipage}
\caption{Throughput, P95 latency, and P99 latency analysis using slides of $1$M edges and windows of various sizes on average.}\label{fig:scale_window_size}
\end{figure*}

\begin{figure*}
\resizebox{0.25\linewidth}{!}{
\includegraphics{figures/exp/scalability-throughput-legend.pdf}}\\
    \begin{minipage}{0.29\linewidth}
        \begin{minipage}{0.32\textwidth}
            \resizebox{\textwidth}{!}{
                \includegraphics{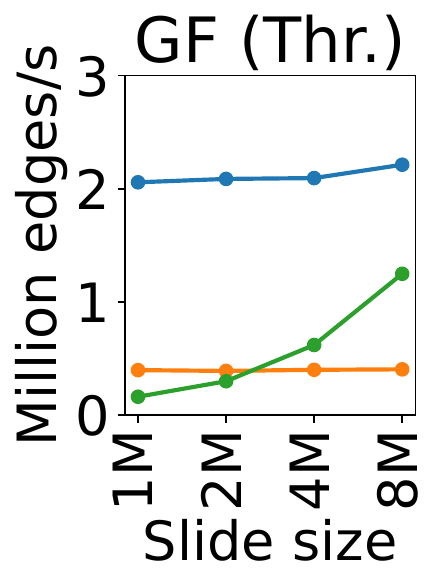}
            }
        \end{minipage}
        \begin{minipage}{0.32\textwidth}
            \resizebox{\textwidth}{!}{
                \includegraphics{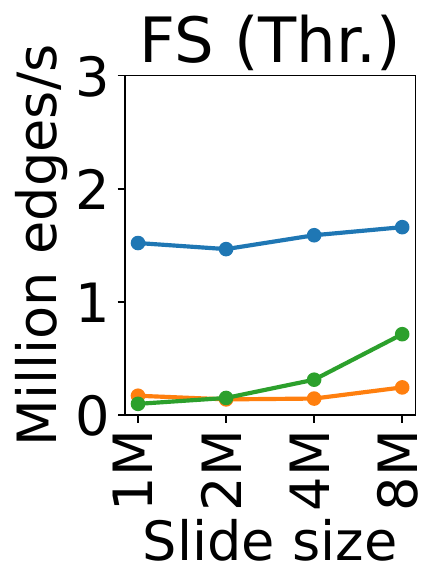}
            }
        \end{minipage}
        \begin{minipage}{0.32\textwidth}
            \resizebox{\textwidth}{!}{
                \includegraphics{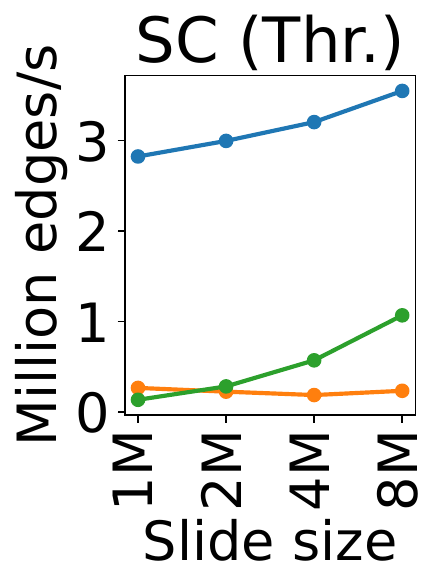}
            }
        \end{minipage}
    \end{minipage}
    \begin{minipage}{0.345\linewidth}
        \begin{minipage}{0.32\textwidth}
            \resizebox{\textwidth}{!}{
    \includegraphics{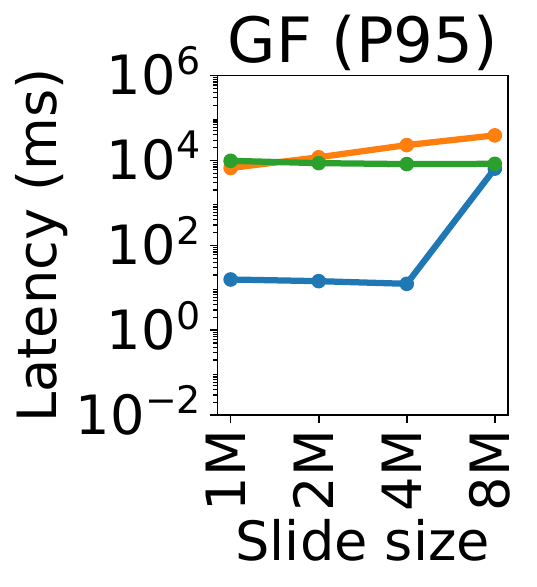}
    }
        \end{minipage}
        \begin{minipage}{0.32\textwidth}
            \resizebox{\textwidth}{!}{
    \includegraphics{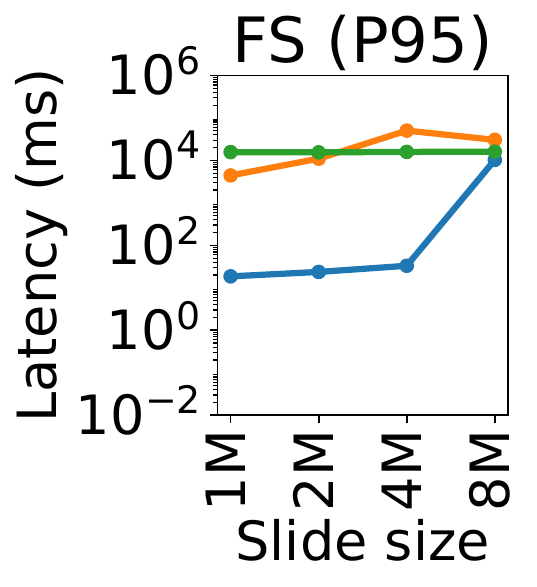}}
        \end{minipage}
        \begin{minipage}{0.32\textwidth}
            \resizebox{\textwidth}{!}{
    \includegraphics{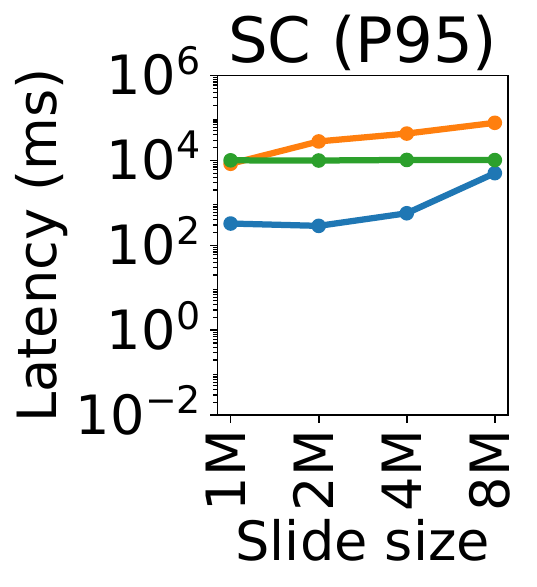}}
        \end{minipage}
    \end{minipage}
    \begin{minipage}{0.345\linewidth}
        \begin{minipage}{0.32\textwidth}
            \resizebox{\textwidth}{!}{
    \includegraphics{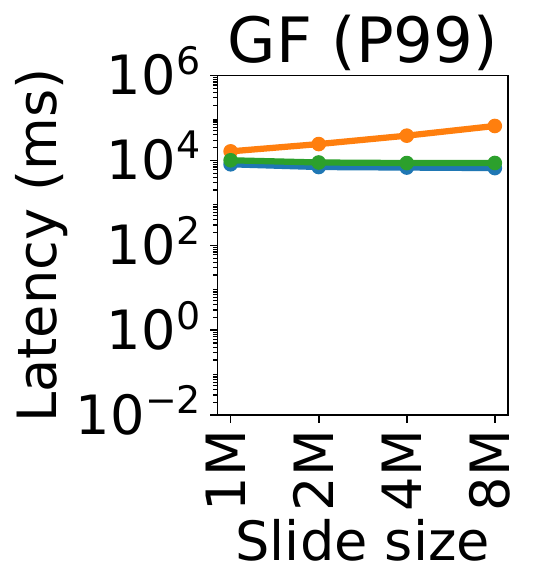}
    }
        \end{minipage}
        \begin{minipage}{0.32\textwidth}
            \resizebox{\textwidth}{!}{
    \includegraphics{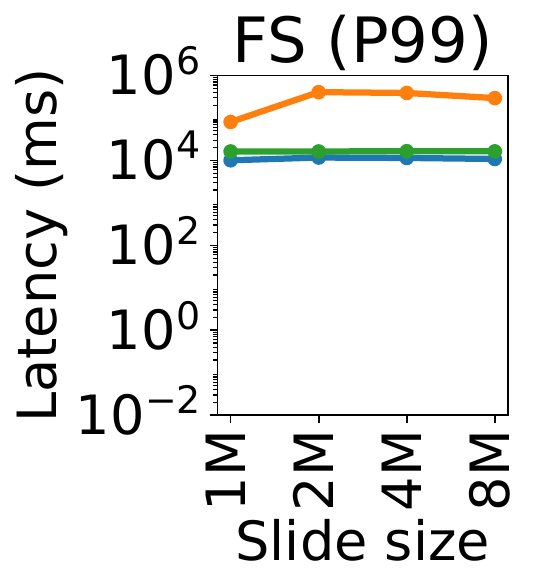}
    }
        \end{minipage}
        \begin{minipage}{0.32\textwidth}
            \resizebox{\textwidth}{!}{
    \includegraphics{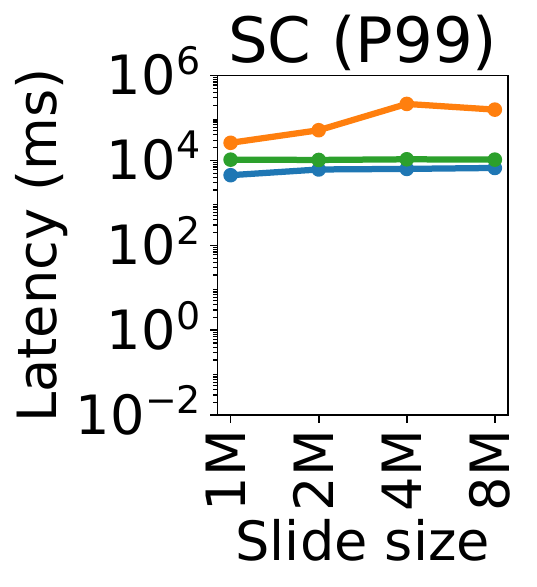}
    }
        \end{minipage}
    \end{minipage}
    \caption{Throughput, P95 latency, and P99 latency analysis using windows of $80$M edges and slides of various sizes on average.}\label{fig:scale_slide_size}
\end{figure*}

\subsection{Throughput and tail latency}\label{sec:performance}
\textit{Throughput}. 
The results of our throughput analysis experiments are displayed in Figure \ref{fig:per_throughput}. 
We conducted throughput experiments for all methods across all datasets.
Notably, the results show that BIC significantly outperforms the compared methods in all datasets. Specifically, BIC improves throughput by up to 14$\times$ over D-Tree, up to 500$\times$ over ET-Tree, up to 1000$\times$ over HDT, and up to  7$\times$ over RWC.
Delving deeper, we find that BIC and RWC exhibit superior throughput over FDC-based approaches, including D-Tree, ET-Tree, and HDT.
The primary reason for this is the considerable overhead associated with deleting expired edges in FDC-based approaches, a process not required by RWC and BIC (detailed in \S \ref{sec:related_works}).
Interestingly, although D-Tree does not boast better time complexities than ET-Tree and HDT, it demonstrates higher throughput. This is largely due to its simplicity in implementation, unlike the complex data structures used in ET-Tree and HDT.
Further distinguishing the performance, the comparison between RWC and BIC reveals a significant throughput difference. 
RWC computes all connected components for each window instance, whereas BIC handles this computation for each chunk. 
In BIC, the computed components in each chunk are then utilized for query processing across 20 window instances in our tested sliding window setup (where the window size is 20$\times$ larger than the slide interval). 
This is a key factor in why BIC shows superior throughput results compared to RWC.

\textit{Tail latency}. 
The results of tail latency analysis experiments are depicted in Figure \ref{fig:per_latency}. 
Overall, BIC significantly outperforms other tested methods in both P99 latency and P95 latency. 
In terms of P99 latency, BIC's improvements are notable: it enhances performance by up to 28$\times$ compared to D-Tree, up to 1500$\times$ over ET-Tree, up to 4500$\times$ over HDT, and up to 2.3$\times$ over RWC. 
BIC's advancements are even more pronounced in P95 latency, where it surpasses other methods by a larger margin: up to 3900$\times$ over D-Tree, up to 100000$\times$ over ET-Tree, up to 400000$\times$ over HDT, and up to 4700$\times$ over RWC.
BIC's performance improvements are due to its specialized computation process. The most compute-intensive scenario in BIC occurs when a streaming edge completes a chunk, necessitating the computation of the chunk's backward buffer. This involves scanning all edges in the chunk to construct the AUFTs (as discussed in \S \ref{sec:aug_backward_uft}). Such intensive computation is only required for the last edge in a chunk, not for every streaming edge. Given that each computed chunk contributes to processing queries in 20 window instances in this experiment, this costly computation significantly impacts P99 latency but is less influential in P95 latency.
Contrastingly, in other methods like RWC and FDC-based approaches, there is no substantial difference between P99 latency and P95 latency. RWC necessitates the computation of all connected components for each window instance before query processing. FDC-based methods face the significant challenge of deleting expired edges, a mandatory step for updating each window instance.

\textit{Discussion on average distance and diameter}.
Average distance and diameter, which are the mean shortest path length and maximum shortest path length in a graph, respectively, affect the performance of FDC methods like ET-Tree, HDT, and D-Tree.
This is because FDC approaches, which are based on spanning trees, necessitate performing a graph traversal (\textit{e.g.}, BFS) to find a replacement edge when a tree edge is deleted in a spanning tree (see \S \ref{sec:related_works} for details). Additionally, average distance and diameter directly impact the efficiency of graph traversal.
In contrast, RWC and BIC, based on Union-Find Trees, are unaffected by these metrics since they don't require graph traversal. 
Due to the computational cost, average distance and diameter are reported in Table \ref{table:datasets} only for the graphs such that their average distance and diameter are available online. 
Generally, these values are low, favoring D-Tree's performance; however, BIC consistently outperforms D-Tree in these scenarios.

\subsection{Impact of window sizes and slide sizes}\label{sec:scalability}
\textit{Window sizes}.
In Figure \ref{fig:scale_window_size}, we present the results of our experiments analyzing throughput and tail latency using windows with a fixed slide interval but varying window sizes. 
Overall, we observe that as the window size increases, the improvement of BIC over D-Tree and RWC remains consistent with the trends reported in Figures \ref{fig:per_throughput} and \ref{fig:per_latency}. Below, we delve into more detailed findings in throughput and latency results, respectively.
Specifically, we observe a significant decrease in RWC's throughput with larger window sizes. This reduction is attributed to RWC's approach of computing connected components for each window instance, where a larger window size directly translates to increased computation time and subsequently reduced throughput.
In contrast, BIC's throughput is less adversely affected by increasing window sizes. This is because, in BIC, the increment in chunk size, which is proportional to the window size, allows for the computed backward buffer to be applied across a broader range of window instances. This efficient utilization helps mitigate the increased computation cost.
D-Tree also experiences a decrease in throughput as the window size grows. This decrease is linked to the expansion of D-Tree's primary data structure (spanning trees), leading to more resource-intensive operations.
Turning to tail latency, we see a clear increase in both P99 latency and P95 latency for RWC as the window size expands. 
This is a direct consequence of the longer time needed to compute connected components in each larger window instance.
For BIC, while the P99 latency similarly increases with window size, an interesting trend is observed in P95 latency. 
We note a significant reduction in P95 latency when the window size shifts from 10M to 20M edges. This phenomenon is explained by the frequency of computing backward buffers in chunks, which is a crucial factor in P95 latency. In smaller windows, this computation happens every 10 instances, but in larger windows, it occurs every 20 instances, effectively reducing the P95 latency. D-Tree's tail latency demonstrates irregular patterns, varying depending on the graph. This inconsistency is anticipated as D-Tree's latency is more sensitive to the slide size rather than the window size.

\begin{figure*}
\resizebox{0.33\linewidth}{!}{
\includegraphics{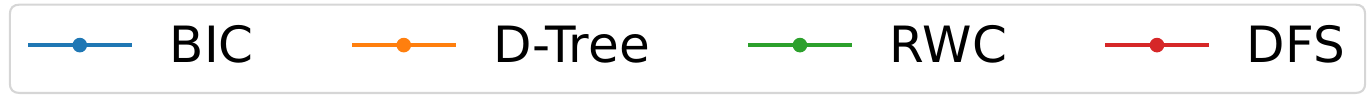}}\\
    \begin{minipage}{0.29\linewidth}
        \begin{minipage}{0.32\textwidth}
            \resizebox{\textwidth}{!}{
    \includegraphics{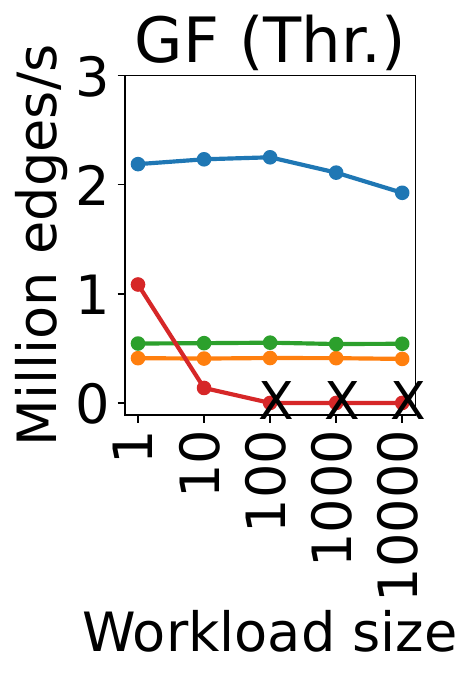}
    }
        \end{minipage}
        \begin{minipage}{0.32\textwidth}
            \resizebox{\textwidth}{!}{
    \includegraphics{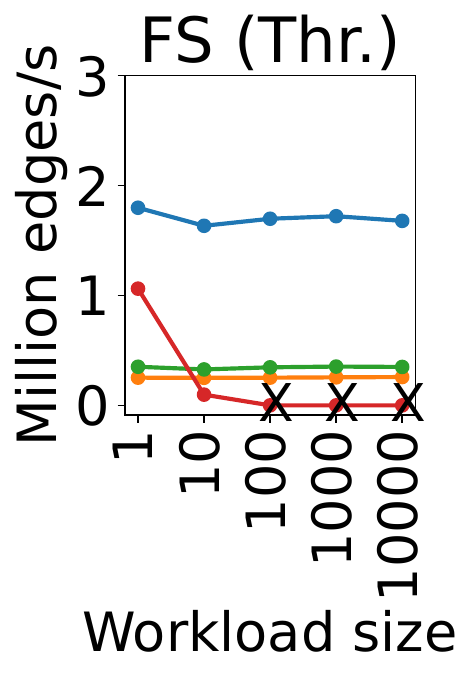}
    }
        \end{minipage}
        \begin{minipage}{0.32\textwidth}
            \resizebox{\textwidth}{!}{
    \includegraphics{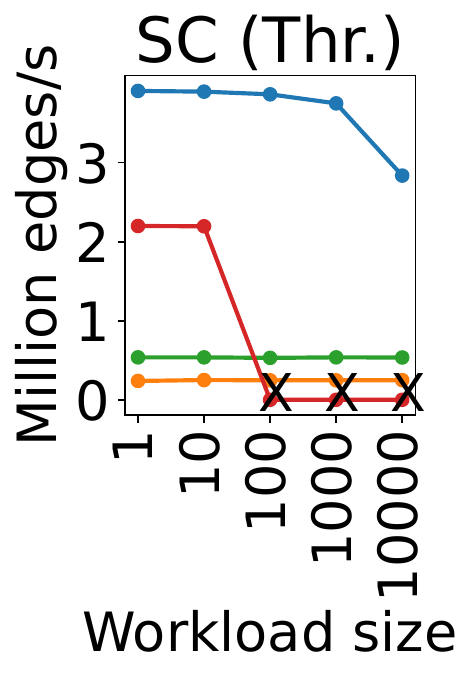}
    }
        \end{minipage}
    \end{minipage}
    \begin{minipage}{0.35\linewidth}
        \begin{minipage}{0.32\textwidth}
            \resizebox{\textwidth}{!}{
    \includegraphics{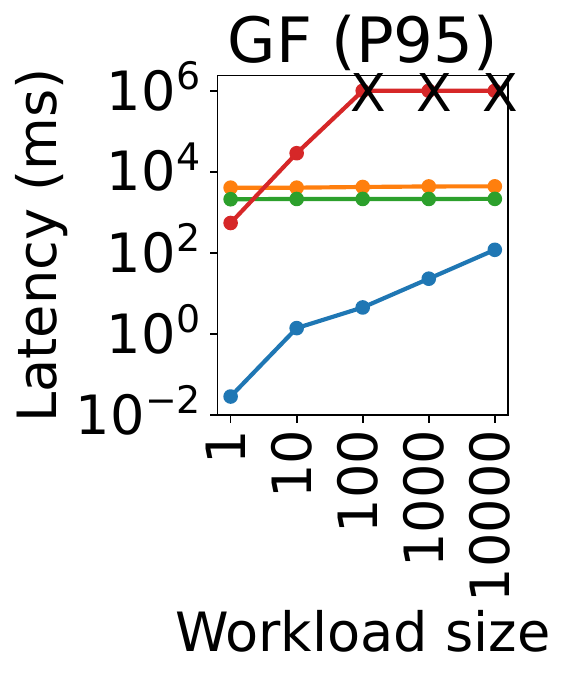}
    }
        \end{minipage}
        \begin{minipage}{0.32\textwidth}
            \resizebox{\textwidth}{!}{
    \includegraphics{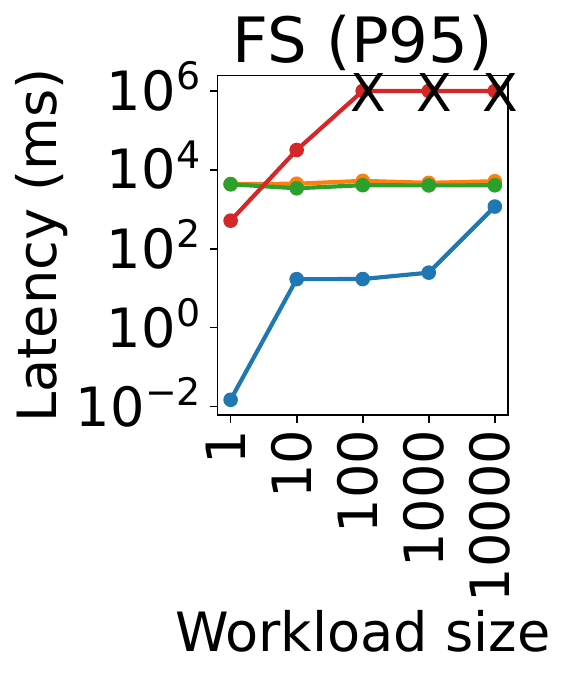}
    }
        \end{minipage}
        \begin{minipage}{0.32\textwidth}
            \resizebox{\textwidth}{!}{
    \includegraphics{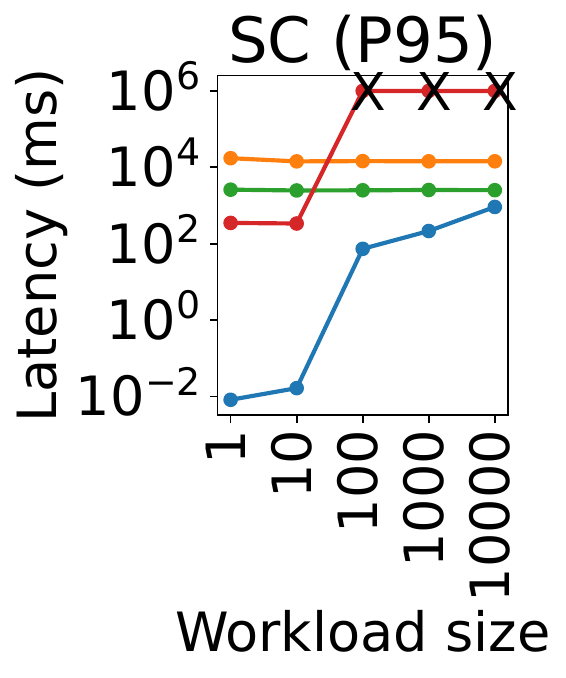}
    }
        \end{minipage}
    \end{minipage}
    \begin{minipage}{0.35\linewidth}
        \begin{minipage}{0.32\textwidth}
            \resizebox{\textwidth}{!}{
    \includegraphics{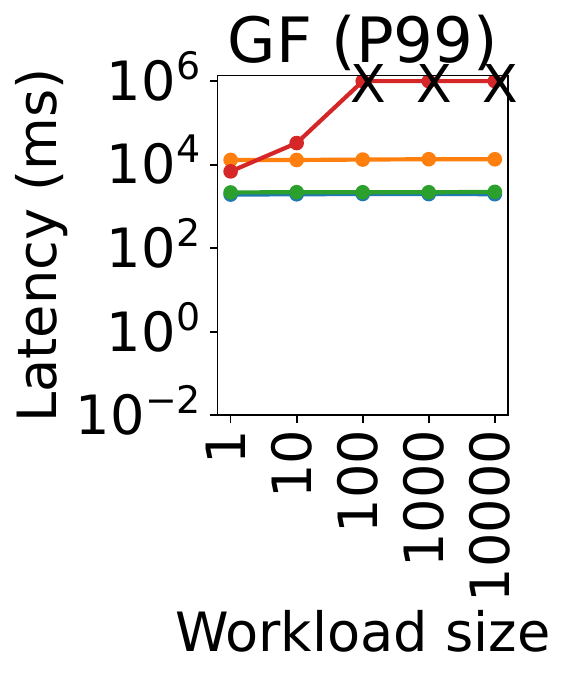}
    }
        \end{minipage}
        \begin{minipage}{0.32\textwidth}
            \resizebox{\textwidth}{!}{
    \includegraphics{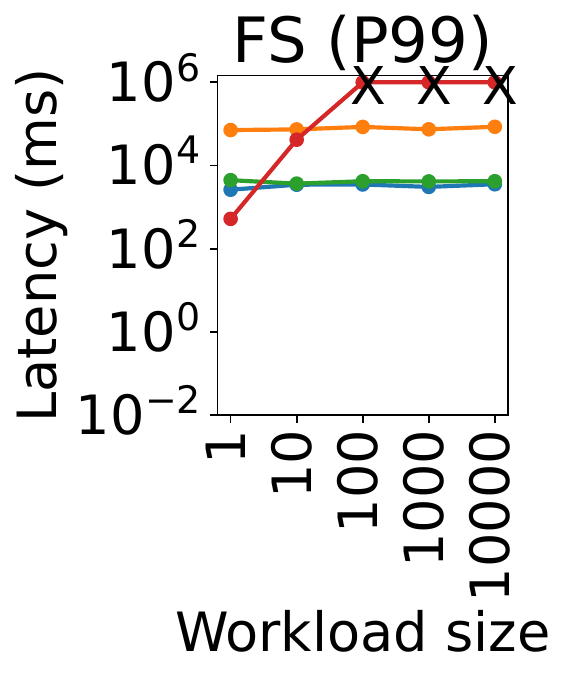}
    }
        \end{minipage}
        \begin{minipage}{0.32\textwidth}
            \resizebox{\textwidth}{!}{
    \includegraphics{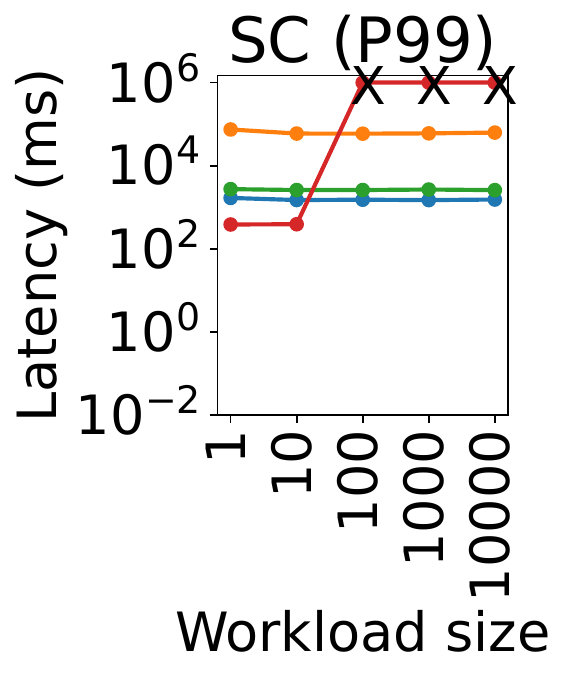}
    }
        \end{minipage}
    \end{minipage}
    \caption{Throughput, P95 latency, and P99 latency analysis using various workload sizes.}\label{fig:scale_workload_size}
\end{figure*}

\begin{figure*}
\begin{minipage}{0.29\linewidth}
    \resizebox{\textwidth}{!}{
        \includegraphics{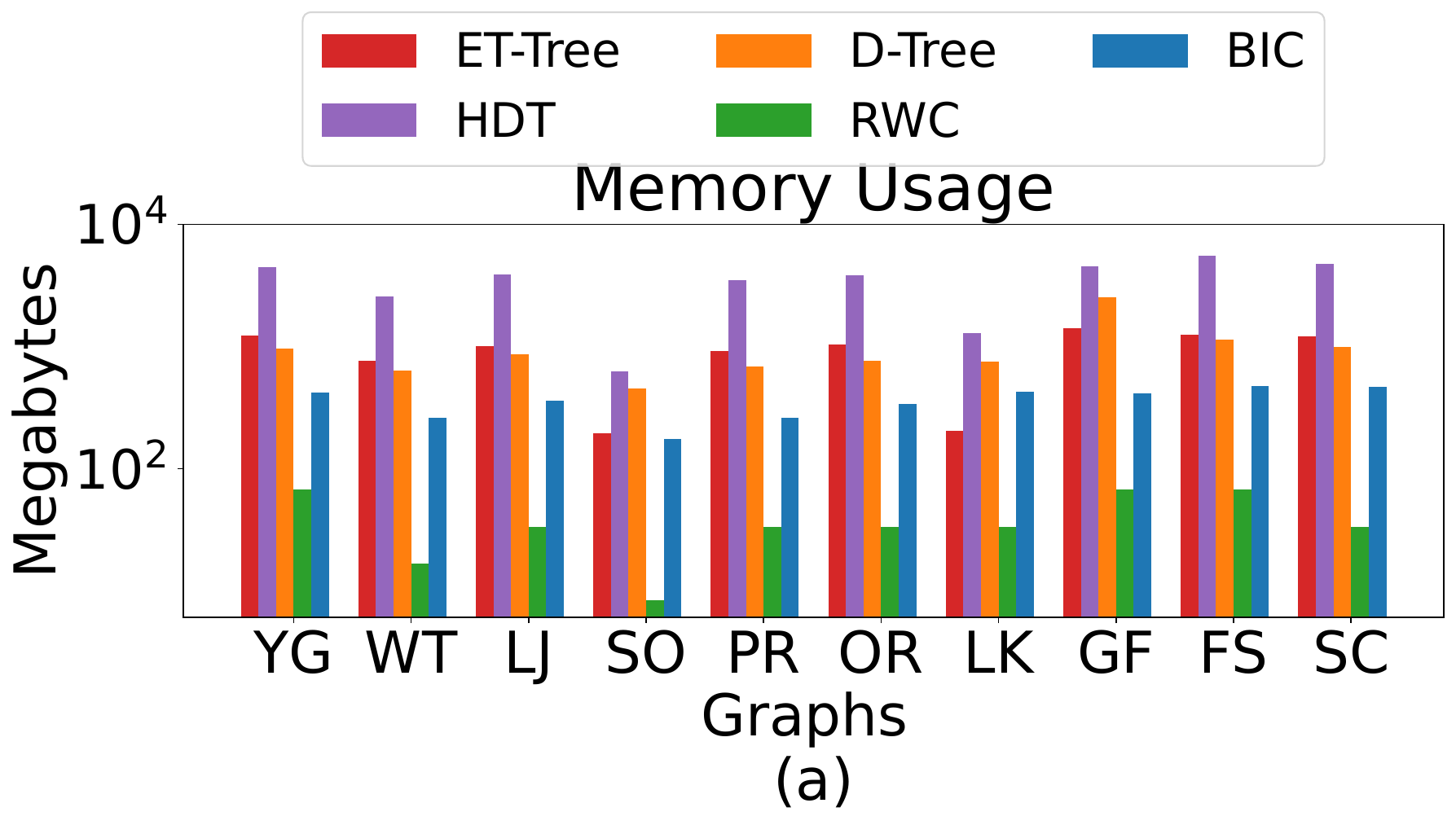}
    }
\end{minipage}
\begin{minipage}{0.69\linewidth}
\begin{minipage}{\linewidth}
\centering
    \resizebox{0.35\linewidth}{!}{
\includegraphics{figures/exp/scalability-throughput-legend.pdf}}     
\end{minipage}\\
    \begin{minipage}{0.5\linewidth}
    \begin{minipage}{0.32\textwidth}
        \resizebox{\textwidth}{!}{
\includegraphics{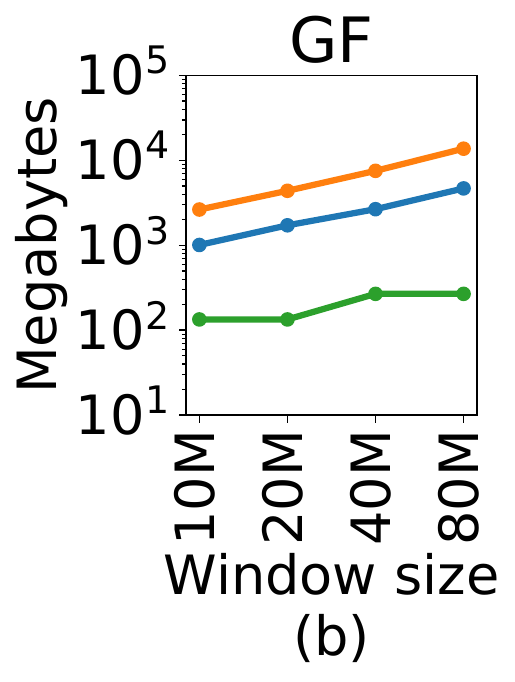}
}
    \end{minipage}
    \begin{minipage}{0.32\textwidth}
        \resizebox{\textwidth}{!}{
\includegraphics{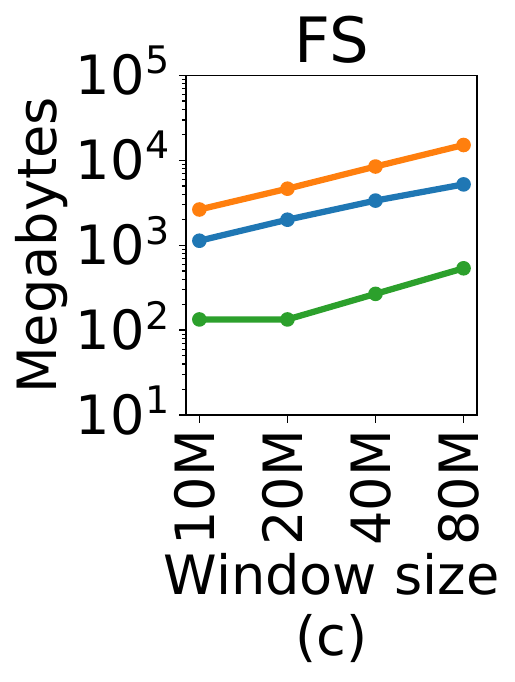}
}
    \end{minipage}
    \begin{minipage}{0.32\textwidth}
        \resizebox{\textwidth}{!}{
\includegraphics{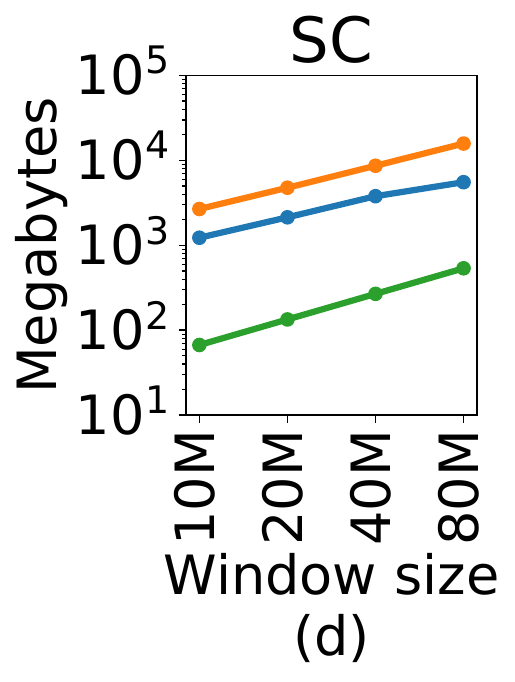}
}
    \end{minipage}
\end{minipage}
    \begin{minipage}{0.5\linewidth}
    \begin{minipage}{0.32\textwidth}
        \resizebox{\textwidth}{!}{
\includegraphics{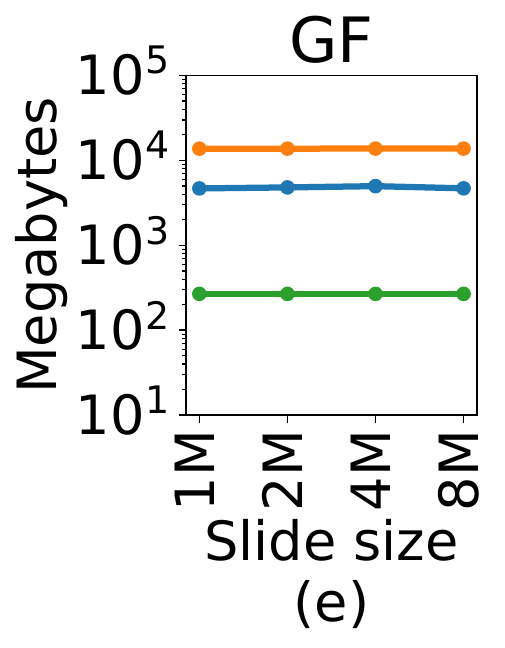}
}
    \end{minipage}
    \begin{minipage}{0.32\textwidth}
        \resizebox{\textwidth}{!}{
\includegraphics{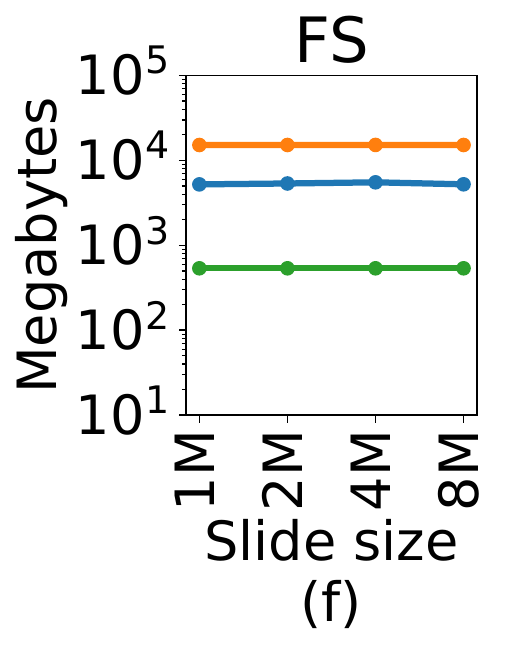}
}
    \end{minipage}
    \begin{minipage}{0.32\textwidth}
        \resizebox{\textwidth}{!}{
\includegraphics{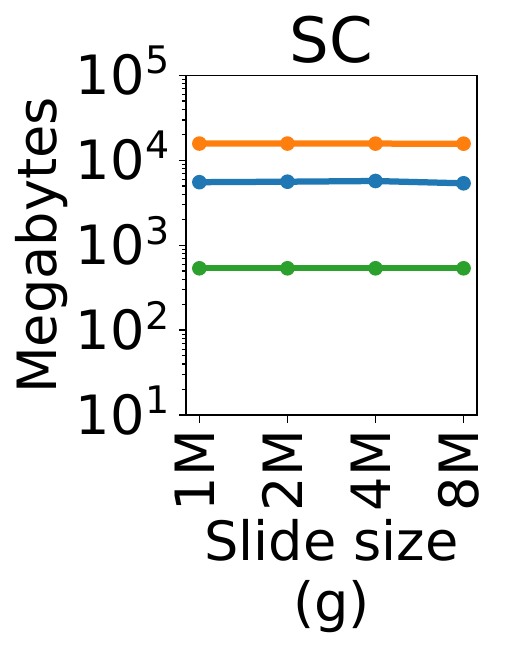}
}
    \end{minipage}
    \\
\end{minipage}
\end{minipage}
\caption{
Memory usage analysis: (a) analyzes per graph; (b)-(d) assess window size impact; (e)-(g) evaluate slide size impact.
}\label{fig:per-memory}
\end{figure*}
 
\textit{Slide sizes}.
In Figure \ref{fig:scale_slide_size}, we present the results from our experiments analyzing throughput and tail latency using windows of a fixed size but with varying slide sizes. Generally, we observe that as the slide size increases, BIC's improvement over D-Tree and RWC in terms of throughput and tail latency remains consistent with the improvements reported in Figures \ref{fig:per_throughput} and \ref{fig:per_latency}. An exception to this trend is noted in the case of BIC's throughput improvement over RWC. We delve into these details below.
There is a noticeable increase in RWC's throughput with larger slide sizes. This is because a larger slide size means less frequent query processing over window instances, thereby reducing the frequency of RWC's major bottleneck — computing the connected component in windows. In extreme cases, where the slide interval equals the window size, window instances become disjoint (aka \textit{tumbling windows}), reducing the need for incremental computation methods like BIC and D-Tree. 
Tumbling windows are generally less challenging than sliding windows.
Both BIC and D-Tree also experience a slight increase in throughput, attributed to the less frequent processing of queries.
Turning to tail latency, we see that the tail latency for RWC remains largely unchanged due to the fixed window size.
BIC shows similar behavior in tail latency, except for P95 latency in cases where the slide size increases from 4M to 8M edges. This pattern mirrors the trend observed in Figure \ref{fig:scale_window_size} when the window size increased from 10M to 20M edges. In cases where the slide size is 8M, the computation time for backward buffers is factored into P95 latency, unlike in scenarios with smaller slide sizes.
Both P95 latency and P99 latency in D-Tree exhibit a noticeable increase with larger slide sizes. This is because D-Tree's tail latency is predominantly influenced by the cost of deleting expired edges, and a larger slide size results in a greater number of edges to delete.

\subsection{Impact of workload size}\label{sec:workload}
In Figure \ref{fig:scale_workload_size}, we present our experimental results on throughput and tail latency across workloads of varying sizes. 
Interestingly, in scenarios with small workloads, such as a single query, neither D-Tree nor RWC exhibit superior performance compared to DFS in terms of throughput and latency. However, BIC significantly outperforms DFS, particularly in terms of P95 latency. This finding suggests that BIC is consistently beneficial for index construction to accelerate query processing. 
The performance of DFS decreases dramatically as the workload size increases, encountering timeouts in large workloads within a 10-hour limit (indicated by `X' in Figure \ref{fig:scale_workload_size}). This is because processing each query requires performing a single DFS.
The influence of workload size on the performance of RWC is minimal. This is because once the connected components for each window instance in RWC are computed, the time taken for query processing becomes insignificant.
D-Tree exhibits similar behaviors as its performance bottleneck is in index updating instead of query processing, and the window and slide size are fixed in this setting.
The impact of workload size on BIC's throughput is slight. For P99 latency, the workload size does not significantly affect BIC since the query processing time is relatively minor compared to the time required for computing backward buffers. P95 latency in BIC tends to increase with larger workloads. This increase is due to the fact that, for each query, BIC might need to search over the maintained BFBG (\S \ref{sec:BFBG}) in the worst-case scenario.

\subsection{Memory usage}\label{sec:memory_udage}
In Figure \ref{fig:per-memory}, we present the memory usage results for all methods, detailing memory usage on each graph in Figure \ref{fig:per-memory}(a), a sliding window with a slide size of $1$M edges and different window sizes in Figures \ref{fig:per-memory}(b)-(d), and a sliding window with a window size of $80$M edges and different slide sizes in Figures \ref{fig:per-memory}(e)-(g).
Index-based methods like ET-Tree, HDT, D-Tree, and BIC maintain an index for streaming edge processing, with their memory efficiency gauged by the median index size across window instances.
RWC, differentiating from index-based approaches, constructs Union-Find Trees per window instance by scanning all edges, thus we capture its memory usage based on the  size of these trees and report the median one of all window instances. 
Generally, RWC consumes the least memory as it stores only vertices. In contrast, FDC methods (ET-Tree, HDT, and D-Tree) require more memory for holding both edges and vertices per window instance.
Among index-based approaches, BIC is the most memory-efficient, utilizing Union-Find Trees for indexing and storing edges per chunk rather than per window instance. Although RWC is less memory-intensive than BIC, it faces latency challenges in constructing Union-Find Trees, as depicted in Figures \ref{fig:per_latency}, \ref{fig:scale_window_size}, and \ref{fig:scale_slide_size}. 
Memory usage escalates with window size due to the increased count of vertices and edges, as shown in Figures \ref{fig:per-memory}(b)-(d), but remains unaffected by slide size changes (Figures \ref{fig:per-memory}(e)-(g)) when the window size is constant.

\section{conclusion}\label{sec:conclusion}
We study index-based query processing in sliding windows over streaming data, focusing specifically on connectivity queries over streaming graphs.
The dynamic nature of sliding windows, characterized by deleting expired edges and inserting new edges, presents inherent challenges in index maintenance.
To address these challenges, we propose BIC, a generic computation model towards high-throughput and low-latency sliding window query processing.
BIC is tailored to circumvent physically performing edge deletions in the maintained indexes, a process typically identified as a performance bottleneck in existing indexes for connectivity queries in dynamic graphs. 
We then propose specialized data structures that synergize with BIC to efficiently process sliding window connectivity.
The results from our comprehensive experimental evaluation highlight the effectiveness of our approach, showcasing up to a 14$\times$ increase in throughput and a reduction in P95 latency by up to 3900$\times$ when compared to state-of-the-art indexes.

We recognize several promising directions for future research. These include the development of optimization and parallelization techniques and the study of using different chunk sizes in BIC, aimed at further enhancing BIC's performance.
Additionally, we envisage the application of the BIC model in broader contexts, extending its capabilities to achieve high throughput and low latency in processing a variety of queries within sliding windows.

\bibliographystyle{ACM-Reference-Format}
\bibliography{publications, bibliography}

\clearpage
\appendix
\section{Proof of Lemma \ref{lemma:snapshot_isolation}}
\label{app:lemma}
\begin{proof}
In the path from $v$ to the root of the UFT containing $v$ in $b_i[j']$, let $(x,y)$ be the first UFTE labeled with a snapshot index that is smaller than $j$. 
Then, each of the edges after $(x,y)$ in the path to the root must be labeled with a snapshot index that is smaller than $j$, because the backward buffer is computed in the backward manner such that UFTEs are inserted with continuously decreasing snapshot indexes. 
\end{proof}

{\small
\begin{algorithm}[h]
\SetAlgoVlined
\SetKwFunction{KwComputingSnapshotIsolation}{computeAndStoreBackwardBuffers}
\SetKwFunction{KwFindRootSnapshotIsolation}{findRootWithSnapshotIsolation}
\SetKwProg{Fn}{procedure}{}{}
\caption{Snapshot isolation of backward buffers}\label{algo:snapshot_isolation}
\Fn{\KwComputingSnapshotIsolation{$c$}}{
$b\gets$ the empty Union-Find data structure\;
    \For{$i\gets |c|-1$ \KwTo $0$ }{    
        \For{$\forall (u,v)\in c[i]$}{            
            $ru\gets $ the root of $u$ in $b$\;
            $rv\gets $ the root of $v$ in $b$\;
            \If{$ru$ is not equal to $rv$}{
                insert $(ru,rv)$ into $b$\;
                label $(ru,rv)$ with $i$ in $b$\;              
            }
        }        
    }
}
\Fn{\KwFindRootSnapshotIsolation{$v,j,b$}}{
    $u\gets$ the parent of $v$ in $b$\;
    $j'\gets$ the label of $(u,v)$ in $b$\;
    \While{$u$ is not null \textbf{and} $j'\geq j $}{
        $v \gets u$\;
        $u \gets $ the parent of $v$ in $b$\;
        $j'\gets$ the label of $(u,v)$ in $b$\;
    }
    \Return $v$\;
}
\end{algorithm}}

\section{Algorithms for storing backward buffers}\label{app:snapshot}
The algorithm for using the snapshot isolation approach to store backward buffers is presented in Algorithm \ref{algo:snapshot_isolation}.
The procedure for computing backward buffers is shown in lines 1-9. Specifically, all edges in a chunk $c$ are scanned in the backward manner, \textit{i.e.}, from $|c|-1$ to $0$ in lines 3-9. For each edge $(u,v)$, if they do not have the same root in the backward buffer, the edge $(u,v)$ is inserted as an edge of the union-find tree in $b$, \textit{i.e.}, line 8. In addition, the edge $(u,v)$ is labeled with the current index number $i$ in $b$, \textit{i.e.}, line 9. 
The procedure for finding the root of a vertex in a backward buffer stored using the snapshot isolation approach is shown in lines 10-17. Specifically, given a vertex $v$, the index number $j$, and a backward buffer $b$, in order to find the root of $v$ in the union-find trees in $b[j]$, the path from $v$ to the root is visited, where every edge in the path is labeled with an index number $j'$ that is not smaller than $j$, \textit{i.e.}, lines 13-16.

\section{Algorithms for computing AUFTs}
\label{app:auft}
The algorithm for computing AUFTs is presented in Algorithm \ref{algo:augmented_backward_buffer}.
Vertices are labeled if they are not contained in the current snapshot of $b$, \textit{i.e.}, lines 18-20.
If a vertex $v$ becomes a root, then the vertex is labeled with an interval, \textit{i.e.}. lines 21-23.
If a root $r$ becomes a child of another root, then the interval of $r$ is updated, \textit{i.e.}, lines 24-27.
Notice that the snapshot index of UFTEs required by the snapshot isolation approach can be computed simultaneously as computing AUFTs, which is shown on line 16 in Algorithm \ref{algo:augmented_backward_buffer}.

{\small
\begin{algorithm}[h]
\SetAlgoVlined
\SetKwFunction{KwComputingAUFTs}{computeBckwardBufferWithAUFTs}
\SetKwFunction{KwLabelVertex}{labelVertex}
\SetKwFunction{KwLabelRoot}{labelRoot}
\SetKwFunction{KwUpdateInterval}{updateInterval}
\SetKwProg{Fn}{procedure}{}{}
\caption{Computing AUFTs in backward buffer $b$}\label{algo:augmented_backward_buffer}
\Fn{\KwComputingAUFTs{$c$}}{
$b\gets$ the empty union-find data structure\;
% $log\gets$ the empty log for $b$\;
    \For{$j\gets |c|-1$ \KwTo $1$ }{    
        \For{$\forall (u,v)\in c[i]$}{      
            \KwLabelVertex{$u,i,b$}\;
            \KwLabelVertex{$v,i,b$}\;
            $ru\gets $ the root of $u$ in $b$\;
            $rv\gets $ the root of $v$ in $b$\;
            \If{$ru$ is not equal to $rv$}{
                \If{$ru$ is linked as a child of $rv$}{
                    \KwLabelRoot{$rv,i$}\;
                    \KwUpdateInterval{$ru,i$}\;
                } \Else{
                    \KwLabelRoot{$ru,i$}\;
                    \KwUpdateInterval{$rv,i$}\;
                }        
                % record $(ru,rv)$ in $log[i]$\; 
                label $(ru,rv)$ in $b$ with snapshot index $j$\;
            }
        }        
    }
    \Return backward buffer $b$\;
}
\Fn{\KwLabelVertex{$v,j,b$}}{
    \If{$v$ is not contained in $b$}{
        label $v$ with $j$\;
    }
}
\Fn{\KwLabelRoot{$r,j$}}{
    \If{$r$ is not labeled with an interval}{
        label $r$ with $[1,j]$;
    }
}
\Fn{\KwUpdateInterval{$v,j$}}{    
    \If{$v$ is labeled with an interval}{
     $[j_s,j_e]\gets$ the interval $v$\;
     set the interval $v$ to $[j+1,j_e]$\;
    }
}
\end{algorithm}
}

{\small
\begin{algorithm}[h]
\SetAlgoVlined
\SetKwFunction{KwComputingEdgesAndIntervals}{computeEdgesAndIntervals}
\SetKwProg{Fn}{procedure}{}{}
\caption{Computing edges and intervals using AUFTs}\label{algo:compute_edges_intervals}
\Fn{\KwComputingEdgesAndIntervals{$b, j, v, v_f$}}{
    $list\gets$ an empty list of edges labeled with intervals\; 
    $l\gets$ the label of $v$ in the AUFT in $b$\;
    $v'_b\gets$ the first vertex in the path from $v$ to the root in the AUFT such that $v'_b$ has interval $[j'_s,j'_e],j'_s\leq l$\;
    $j'_e\gets min(l,j'_e)$\;
    \If{$v'_b$ is the root}{   
        add $(v'_b, v_f)$ labeled with $[j,j'_e]$ into $list$\;
    } \Else{        
        add $(v'_b, v_f)$ labeled with $[j'_s,j'_e]$ into $list$\;
        $v''_b\gets$ the parent of $v'_b$\;
        $temp\gets j'_s-1$\;
        \While{the parent of $v''_b$ is not empty}{             
            $[j''_s,j''_e]\gets $ the interval of $v''_b$ in $AUFT$\;   
            add $(v''_b, v_f)$ labeled with $[j''_s,temp]$ into $list$\;                   
            $v''_b\gets$ the parent of $v''_b$\;
            $temp \gets j''_s-1$\;                                        
        }        
        add $(v''_b, v_f)$ labeled  with $[j, temp]$ into $list$\;
    }
    \Return $list$\;
}
\end{algorithm}
}

We discuss how to compute the intervals of edges in a BFBG using AUFTs.
Consider an inter-vertex $v$ and its root $v_f$ in forward buffer $f_{i+1}[j-1]$.
In order to record the connection between $b_i[j]$ and $f_{i+1}[j-1]$ via $v$, we need to compute the root $v_b$ of $v$ in $b_i[j]$ and insert edge $(v_b,v_f)$ with an interval into the BFBG.
In addition, as discussed in \S \ref{sec:BFBG}, additional edges $(v'_b,v_f)$ with intervals might need to be inserted if $v$ has root $v'_b$ that is different from $v_b$ in $b_i[j']$, $j'>j$. 
With the AUFT containing $v$ in $b_i[j]$, the vertex label $l$ of $v$ and the path from $v$ to root $v_b$ are retrieved to compute all the possible roots in $b_i[j]$ and each $b_i[j']$, $j'>j$.
Specifically, the path from $v$ to $v_b$ is traversed and each vertex in the path is checked.
Assuming $v'_b$ is the first vertex in the path, such that $v'_b$ has an interval, which is $[j'_s,j'_e]$.
If $l < j'_s$, no edge will be inserted into the BFBG for $v'_b$ because in this case, $v$ is not in the AUFT rooted at $v'_b$.
Otherwise, edge $(v'_b,v_f)$ with an interval will be inserted into the BFBG. 
The interval is computed according to the relationship between vertex label $l$ and interval $[j'_s,j'_e]$. 
If $j'_e< l$, then the interval of $(v'_b,v_f)$ will be $[j'_s,j'_e]$. 
If $j'_s\leq l\leq j'_e$, then $(v'_b,v_f)$ will be labeled with interval $[j'_s,l]$.
For the next vertex $v''_b$ that has interval $[j''_s,j''_e]$, edge $(v''_b, v_f)$ with interval $[j''_s,j'_s-1]$ is inserted into the BFBG.
The last vertex in the path will be the root $v_b$ of $v$ in $b_i[j]$. 
The edge  $(v_b,v_f)$ labeled with $[j, j''_s -1]$ will be inserted into BFBG, where $j$ is the current snapshot index, and $j''_s$ is the first endpoint of the interval for the second vertex from the last in the path from $v$ to $v_b$.

The algorithm for computing edges labeled with intervals is shown in Algorithm \ref{algo:compute_edges_intervals}. The input of the algorithm is the current snapshot of backward buffer $b$, the current snapshot index $j$ in $b$, the inter-vertex $v$, and the root $v_f$ of $v$ in forward buffer $f[j-1]$.
The algorithm computes a list of edges labeled with intervals, which will be inserted into BFBG.
If the root $v_b$ of $v$ in the path from $v$ to $v_b$ is the only vertex that is labeled with a non-empty interval, then the edge $(v_b, v_f)$ is inserted, \textit{i.e.}, lines $6-7$.
Otherwise, for each vertex that is labeled with a non-empty interval, a corresponding edge is inserted, \textit{i.e.}, lines $9$-$18$.

\section{Computing BFBGs}\label{app:BFBG}
The algorithm for computing a BFBG is presented in Algorithm \ref{algo:compute_bfbg}.
Given a streaming edge $(u,v)$, the algorithm processes vertices $u$ and $v$ respectively and then inserts edges with intervals into the current BFBG. 
Consider processing $v$. If $v$ is an inter-vertex, \textit{i.e.}, $v$ exists in both the corresponding backward and forward buffers, the roots of $v$ in the forward and backward buffers are computed, \textit{i.e.}, lines 3-7. 
Once the root of $v$ in the forward buffer is obtained, the algorithm calls Algorithm \ref{algo:compute_edges_intervals} to obtain the list of edges and intervals that need to be inserted into the current BFBG, \textit{i.e.}, line 5, and then perform such insertion operations, \textit{i.e.}, lines 6-7.

{\small
\begin{algorithm}[h]
\SetAlgoVlined
\SetKwFunction{KwComputingBFBG}{processVertex}
\SetKwProg{Fn}{procedure}{}{}
\caption{Computing BFBGs}\label{algo:compute_bfbg}
    \Fn{\KwComputingBFBG{$v$}}{
        $(b_i[j], f_{i+1}[j-1])\gets$ the current elements in the backward and forward buffers, respectively\;
        \If{$v\in b_i[j]$ \textbf{and} $v\in f_{i+1}[j-1]$}{
            $v_f\gets$  the root of $v$ in $f_{i+1}[j-1]$\;
            $list\gets$  \texttt{computeEdgesAndIntervals}($b,j,v,v_f$)\;
            \For{$\forall ((v_b,v_f),[j_s,j_e]) \in list$}{
                insert $(v_b,v_f)$ into the current BFBG and label $(v_b,v_f)$ with $[j_s,j_e]$\;
            }
        }
    }
\end{algorithm}
}

\section{Query processing}\label{app:query_processing}
The algorithm for query processing is presented in Algorithm \ref{algo:query_processing}.
For a query $Q_c(u,v)$, the algorithm first checks whether $u$ and $v$ are connected in either the corresponding forward buffer or the corresponding backward buffer, which is computed by checking whether they have the same roots in each buffer, \textit{i.e.}, lines 3-8. If such connectivity information can be determined, query result $True$ can be immediately returned. This corresponds to the intra-buffer checking in the query processing.
Otherwise, the intra-buffer checking is performed. Specifically, the BFBG represents the connectivity information between $b_i[j]$ and $ f_{i+1}[j-1]$ is first retrieved, \textit{i.e.}, line 9.
Then, a root $r_u$ of $u$ in either the forward or the backward buffer is retrieved, \textit{i.e.}, lines 11-14, where the condition for checking whether the root is null is necessary because $u$ may not exist in both buffers. Similarly, a non-null root $r_v$ of $v$ is also retrieved. Finally, $Q_c(u,v)$ is determined by whether $r_u$ and $r_v$ are connected on the BFBG, \textit{i.e.}, lines 19-22, which can be simply implemented by performing a BFS on the BFBG with a constraint that only edges labeled with an interval subsuming $j$ is visited.

{\small
\begin{algorithm}[h]
\SetAlgoVlined
\SetKwFunction{KwQueryProcessing}{Query}
\SetKwProg{Fn}{procedure}{}{}
\caption{Query processing}\label{algo:query_processing}
    \Fn{\KwQueryProcessing{$u,v$}}{
        $(b_i[j], f_{i+1}[j-1])\gets$ the current elements in the backward and forward buffers, respectively\;
        $f_u, f_v\gets$ the roots of $u$ and $v$ in $f_{i+1}[j-1]$, respectively\;
        \If{$f_u = f_v$}{
            \Return True\;
        }
        $b_u, b_v\gets$ the roots of $u$ and $v$ in $b_{i}[j]$, respectively\;
        \If{$b_u = b_v$}{
            \Return True\;
        }
        $BFBG \gets$ the current BFBG for $(b_i[j], f_{i+1}[j-1])$\;
        initialize two vertices $r_u$ and $r_v$\;
        \If{$f_u$ is not null}{
            $r_u\gets f_u$\;
        }\Else{
            $r_u\gets b_u$\;
        }
        \If{$f_v$ is not null}{
            $r_v\gets f_v$\;
        }\Else{
            $r_v\gets b_v$\;
        }
        \If{$r_v$ \textbf{and} $r_u$ are connected in BFBG}{
            \Return True\;
        }\Else{
            \Return False\;
        }
    }
\end{algorithm}
}

\end{document}